\documentstyle[epsf,supertb]{l-aa.pr}

\def\hi{\relax \ifmmode {\rm H\,{\sc i}}\else H\,{\sc i}\fi}
\def\hii{\relax \ifmmode {\rm H\,{\sc ii}}\else H\,{\sc ii}\fi}
\def\rq{$r^{1/4} $ }
\def\rh{$r^{1/2} $ }
\def\muo{\relax \ifmmode \mu_0\else $\mu_0$\fi}
\def\mue{\relax \ifmmode \mu_{\rm e}\else $\mu_{\rm e}$\fi}
\def\re{\relax \ifmmode r_{\rm e}\else $r_{\rm e}$\fi}
\def\magarc{mag arcsec$^{-2}$}
\def\boundboxo{\epsfbox[30 190 538 540]}
\def\boundboxt{\epsfbox[10 190 518 540]}
\def\abst{
 In this paper I present a new two-dimensional decomposition technique,
which models the surface photometry of a galaxy with an exponential
light profile for both bulge and disk and, when necessary, with a
Freeman bar. The new technique was tested for systematic errors on both
artificial and real data and compared with widely used one-dimensional
decomposition techniques, where the luminosity profile of the galaxy is
used.  The comparisons indicate that a decomposition of the two-dimensional
image of the galaxy with an exponential light profile for both bulge
and disk yields the most reproducible and representative bulge and disk
parameters. 

An extensive error analysis was made to determine the reliability of
the model parameters.  If the model with an exponential bulge profile
is a reasonable description of a galaxy, the maximum errors in the
derived model parameters are of order 20\%.  The uncertainties in the
model parameters will increase, if the exponential bulge function is
replaced by other often used bulge functions as the de Vaucouleurs law.

All decomposition methods were applied to the optical and near-infrared
data set presented by de Jong \& van der Kruit (\cite{deJ1}), which
comprises 86 galaxies in six passbands. 

\keywords{Methods: data analysis -- Surveys -- Galaxies: fundamental parameters
 -- Galaxies: photometry -- Galaxies: spiral -- Galaxies: structure}
}

\begin{document}

 \thesaurus{03(03.13.2; 04.19.1; 11.06.2; 11.16.1; 11.19.2; 11.19.6)}
 \title{
Near-infrared and optical broadband surface photometry of 86 face-on
disk dominated galaxies. 
 \thanks{Based on observations with the Jacobus Kapteyn Telescope
operated by the Royal Greenwich Observatory at the Observatorio del
Roque de los Muchachos of the Instituto de Astrof\'\i sica de Canarias
with financial support from the PPARC (UK) and NWO (NL) and with the UK
Infrared Telescope at Mauna Kea operated by the Royal Observatory
Edinburgh with financial support of the PPARC. \newline
\indent Tables \ref{help} are also available in electronic form: see the
Editorial in A\&A 1994 Vol.~280(3) p.~E1. 
 }
} 
 \subtitle{II. A two-dimensional method to determine bulge and disk parameters.}
 \author{Roelof S.~de Jong}
 \offprints{R.S.~de Jong, University of Durham, Dept.~of
Physics, South Road, Durham, DH1 3LE, United Kingdom, e-mail:
R.S.deJong@Durham.ac.uk} 
 \institute{Kapteyn Astronomical Institute, P.O.Box 800, NL-9700 AV
Groningen, The Netherlands}
 \date{received May 19, accepted October 9 1995}
 \maketitle
 \markboth{R.S.~de Jong}{Near-IR and optical observations of 
 86 spirals. II Determination of bulge and disk parameters}


\section{Introduction}
\label{Intro}

The light distribution of disk dominated galaxies is often decomposed
into a bulge and a disk component which are assumed to be physically
and dynamically distinct.  The disk component is flat and governed by
rotational dynamics.  The spherical bulge component, though mainly
rotationally supported against gravity, is dynamically a much hotter
system than the disk.  Whether this separation is real is hard to say;
it is likely that a dynamical interplay exists between the different
components in the dense inner regions of galaxies. Separating both
components using only the surface photometry of a galaxy has been a
longstanding problem.

When we assume that there are distinct components, the parameters
describing the light distribution of these components are of
fundamental importance.  They reveal the common properties among
galaxies and especially in combination with dynamics and (chemical)
content they are tracers of galaxy formation and evolution.  Two
well-known relations with fundamental component parameters are the
constancy of central surface brightness among disks of spiral galaxies
(Freeman \cite{Freeman}) and the link between bulge-to-disk (B/D) ratio
and the Hubble classification sequence.

Many different decomposition techniques can be found in the literature
(for reviews see e.g.~Simien \cite{Sim89}, Capaccioli \& Caon
\cite{CapCao92}).  For most of the methods one postulates some
mathematical functions describing the shape of the different
components, after which the sum of the components functions is fitted
to the observed light distribution.  Decomposition techniques differ in
both the assumed mathematical functions as well as in the applied
fitting algorithms. 

The most frequently used function describing the radial surface
brightness profile of the disks of spiral galaxies is the exponential
function
 \begin{equation} 
\Sigma_{\rm disk}(r) = \Sigma_0 {\rm e}^{-r/h}
\label{explaw}
 \end{equation} 
  or in magnitudes
 \begin{equation} 
\mu_{\rm disk}(r) = \mu_0 + 1.086 r/h
\label{explawmag}
 \end{equation} 
 with \muo\ ($\Sigma_0$) being the central surface brightness
(luminosity) and $h$ the scalelength of the exponential disk.  The
exponential function is not always a good description of the disk light
profile.  After the bulge light is subtracted from the luminosity
profile, the disk profiles of spiral galaxies sometimes show a deficit
of light in their inner regions (Type~II profiles,
Freeman~\cite{Freeman}).  An improved disk model in the form of a
modified exponential profile (Kormendy \cite{Kor77}) is therefore
sometimes fitted to these Type~II profiles.  Furthermore, it should be
noted that the exponential model profile might need a truncation. 
Especially in edge-on galaxies a sudden decrease in light has often been
detected at large radii (van der Kruit \cite{Kruit79}). 


Compared to the disk profile there is less consensus on the
mathematical function to be used for the bulge light profile, because
only the central bright part of the bulge can be seen in most spiral
galaxies.  Away from the center the bulge light is hidden underneath
the disk light for a face-on galaxy and therefore only the central
region can be used to determine the shape of the bulge luminosity
profile.  The most widely used bulge function is the \rq law (de
Vaucouleurs~\cite{deV48}).  This function generally gives a good
description of the light distribution of elliptical galaxies and was
first used for spiral bulges, in combination with an exponential disk,
by de Vaucouleurs (\cite{deV59}).
 The motivation to use the same fitting function for elliptical galaxies
and bulges is an assumed evolutionary or at least structural sequence
from ellipticals to spirals (or vice versa).  The shape of elliptical
luminosity profiles are not completely undisputed and therefore all
light profile functions proposed for ellipticals (e.g.\ Hubble profiles
(\cite{Hub30}), King profiles (\cite{King66}), Jaffe profiles
(\cite{Jaffe83}) or generalized exponentials (Caon et al.~\cite{Caon93};
D'Onofrio et al.~\cite{DOn94})) could in principle also be used for
bulges.  Kormendy (\cite{Kor77}) showed that the parameters of the
Hubble, King and de Vaucouleurs profiles describe the same physical
quantities for ellipticals. 


One should be careful in making the link between elliptical galaxies
and bulges of spiral galaxies.  The presence of the disk will influence
the dynamics of the bulge, especially for late-type systems with a low
B/D ratio.  Bulges are for a considerable fraction rotationally
supported against gravity and not largely pressure supported as are
elliptical galaxies (Kormendy \& Illingworth \cite{KorIll82}; Kormendy
\cite{Kor93} and references therein).  Also deviations from the \rq law
have been observed in bulges of edge-on spirals (Frankston \& Schild
\cite{FraSch76}, Kormendy \& Bruzual \cite{KorBru78}, Burstein
\cite{Bur79}, Jensen \& Thuan~\cite{JenThu79}, Shaw \&
Gilmore~\cite{ShaGil89}, Wainscoat et al.~\cite{Wai89}, Kent et
al.~\cite{Kent91}).  These deviations have motivated Frankston \&
Schild (\cite{FraSch76}), Kent et al.~(\cite{Kent91}) and Andredakis \&
Sanders (\cite{AndSan94}) to propose exponential functions for bulge
profiles. 

Decomposition techniques using the change in ellipticities and position
angles of the isophotes (Kent \cite{Kent86}) have the advantage that
they do not have to assume fitting functions.  They can work perfectly
well assuming that spiral galaxies are only made of ellipsoids/tori with
changing inclinations.  But as real galaxies contain bars, spiral
arms and dust lanes these methods will have systematic errors.  They
only work reasonably well on systems with a high inclination, where the
difference in the flattening of the disk and the bulge component is
easily measurable. 


Decompositions using all the pixels in the full image instead of the
one-dimensional (1D) profile of galaxies take little precedence in the
literature. The main reason for this being the lack of computer power
and an additional reason being the difficulties due to the presence of
bars and spiral arms. These are conveniently averaged out in 1D
profiles. Two-dimensional (2D) fitting techniques have been applied
before on small samples of elliptical galaxies and S0's (Capaccioli et
al.~\cite{CapHel87}; Simien \& Michard~\cite{SimMic90}; Scorza \&
Bender~\cite{ScoBen90}), but only once before on a large set of spiral
galaxies (Byun~\cite{Byun92}). Two-dimensional fitting has the same
advantage as Kent's method (Kent~\cite{Kent86}) in that one uses the
difference in projected ellipticities of disk and bulge. The sample
used here was selected to be face-on and the differences in
ellipticities are expected to be small. Still the 2D fitting technique
is applied, because it has the advantage to 1D fitting that
non-axisymmetric components can be fitted as well. The method applied
here has a non-axisymmetric component in the form of a bar, which will
improve the fitting results of the disk and especially the bulge
component. 

The bulge and disk parameters determined by the 2D method will be used
in subsequent papers to determine the relationships among the structural
parameters of galaxies.  To assess the reliability of these relationships a
thorough error analysis is needed.  In this paper I will discuss several
sources of error, most notably the effects of 1D versus 2D fitting, the
uncertainty in the shape of disk and bulge profiles, influences of
measurement errors and the effect of different radial weighting
functions. 

The structure of this paper is as follows. The observational data are
briefly described in Sect.~\ref{data2}.  The 2D fitting technique is
explained and tested in Sect.~\ref{decomp2D}.  In
Sect.~\ref{compsect} the 2D fitting technique is compared with
several 1D fitting techniques.  The results of all tests and
comparisons are used in the error discussion in Sect.~\ref{discus2}. 
The main conclusions are summarized in Sect.~\ref{concl2}.

\section{The data}
\label{data2}

In order to examine the parameters describing the global structure of
spiral galaxies, 86 systems were observed in the $B, V, R, I, H$ and $K$
passbands.  A full description of the observations and data reduction
can be found in de Jong \& van der Kruit (\cite{deJ1}, hereafter
Paper~I), which will be repeated only very briefly here.  The galaxies
in this statistically complete sample of undisturbed spirals were
selected from the UGC (Nilson \cite{Nilson}) to have red diameters of at
least two minutes of arc and minor over major axis ratios larger than
0.625.  The galaxies were imaged along the major axis with the 1m
Jacobus Kapteyn Telescope on La Palma in the $B, V, R$ and $I$ passbands
and with the United Kingdom Infra-Red Telescope on Hawaii in the $H$ and
$K$ passbands.  Standard reduction techniques (bias subtraction,
flatfielding by twilight flatfields, calibration by standard stars) were
used to produce calibrated images.  The sky brightness was determined
outside the galaxy in areas free of stars and its uncertainty (mainly
due to flatfield limitations) constitutes one of the main sources of
error in the derived parameters. 

The ellipticity and position angle (PA) of each galaxy were determined
at an outer isophote.  The radial surface brightness profiles were
determined by calculating the average surface brightness on elliptical
rings of increasing radii using the determined ellipticity and PA. 
Internal and external comparisons showed that the derived parameters
were well within the estimated errors.  These estimated errors are
included in the analysis discussed here. 

\section{Two-dimensional decomposition}
\label{decomp2D}

In this section the 2D fitting technique is described. The motivation
for using the 2D method was the large number of galaxies in the sample
with a pronounced bar, which can not be fitted in 1D models. The
different model components are described and the fitting procedure
followed is explained in some detail. The fitting technique was tested
on both artificial and real data and is shown to be very accurate in
most realistic cases. Finally, the results for the data set are
presented for the $B$ and the $K$ passband.

\subsection{Advantages of two-dimensional fitting}

In the literature one encounters mainly the use of 1D profiles to
perform bulge/disk decompositions. The extraction of the profiles
improves the signal-to-noise, but the non-axisymmetric information
present in the image is lost.  Bars can have considerable influence on
profiles (see for example in Paper~I UGC\,89, UGC\,6536, UGC\,7523,
UGC\,7594, UGC\,8865 and UGC\,12776).  These features will make 1D
bulge/disk decompositions incorrect, even if the fits seem correct and
the $\chi^2$ values are low.  One-dimensional models fitted to
azimuthally averaged profiles can never include a bar.

A considerable fraction of spiral galaxies are barred.  Of the 86
galaxies of our sample only 13 were classified as non-barred according
the RC3 (de Vaucouleurs et al.~\cite{rc3}) and 12 galaxies had no bar
classification.  Therefore fitting bars is desirable, especially if one
considers that bars are more pronounced in the near-IR (Block \&
Wainscoat~\cite{BloWai91}). This can only be done by fitting the full
2D images.  An additional advantage of the 2D technique is that the
difference in the flattening of the bulge and the disk is also used for
the few systems in the sample with higher inclination. 

The technique of fitting models to the full images of spiral galaxies
has little precedence in the literature.  Shaw \& Gilmore
(\cite{ShaGil89}) fitted 2D models to two edge-on galaxies, the
configuration where disk and bulge are the most distinct. 
Byun~(\cite{Byun92}) used a data set of 1355 $I$ passband images of
galaxies with inclinations larger than $i>40^{\rm o}$.  From tests on
artificial data, Byun showed that the 2D method was better in
reproducing the model input parameters than the 1D method.  Neither
Shaw \& Gilmore nor Byun included a bar in their fits. 

\subsection{The model components}

The 2D model consists of two or three components.  First of all, a
spherically symmetrical bulge with an exponential radial light
distribution.  The use of an exponential bulge profile instead of the more
widely used \rq law profile (de Vaucouleurs~\cite{deV48}) was motivated
by the work of Andredakis \& Sanders~(\cite{AndSan94}) and the work
presented in Sect.~\ref{compsect}. Bulge parameters are normally expressed
in effective parameters which translates the exponential law into
 \begin{equation}
\Sigma_{\rm bulge}(r) = \Sigma_{\rm e} {\rm e}^{-1.679({r/r_{\rm e}}-1)},
\label{expbul}
 \end{equation}
 where the effective radius (\re) encloses half the total luminosity and
$\Sigma_{\rm e}$ is the surface brightness (\mue\ in mag) at this
radius.  In the cartesian coordinate system of the CCD image $r$
should be read as $\sqrt{x^2+y^2}$, with the center of the coordinate
system at the galaxy center. 

The second component, the disk, is described by Eq.~(\ref{explaw}) and
has the usual two free parameters of an exponential light distribution
($\Sigma_0$ and $h$).  Due to inclination though, the disk has two
additional parameters: minor over major axis ratio ($b/a$) and position
angle (PA) and $r$ should be read as 
 $\sqrt{ ([x\cos({\rm PA})+y\sin({\rm PA})]\frac{b}{a})^2 +(-x\sin({\rm PA})+y\cos({\rm
PA}))^2}$ in the carthesian coordinate system.  
 Fitting with $b/a$ and PA as free parameters was tried, but it turned
out that the fitting routine often adjusted the $b/a$ and PA in such
a way that spiral arms and bars were modeled, instead of the global disk
properties.  Therefore I decided to keep $b/a$ and PA fixed to the
values determined at the outer isophotes (the same values that were used
for extracting the radial profiles). 

When the galaxy image showed a clear bar and when this bar resulted in
an identifiable feature in the luminosity profile, a bar was added as a
third component to the model.  A Freeman bar (\cite{Fre66}) was used,
which is one of the few available analytic 2D descriptions for the
luminosity of a bar
 \begin{equation}
\Sigma_{\rm bar}(x,y) = \Sigma_{\rm 0,bar} \sqrt{1-(x/a_{\rm bar})^2-(y/b_{\rm bar})^2} ,
 \end{equation}
 where the free parameters are $\Sigma_{\rm 0,bar}$, the bar central
surface brightness, and $a_{\rm bar}$ and $b_{\rm bar}$, the semi major
and minor axis of the bar respectively.  Such a bar has elliptical
isophotes and has its position angle  (PA$_{\rm bar}$) as additional
free parameter.  No inclination dependent corrections were applied to
the bar profiles, since the studied galaxies are not inclined very
much.

Observations of galaxies are distorted by seeing. The model light
distributions have to be corrected for this effect. Seeing is only
important at the center of the galaxy where the light distribution
strongly peaks. The disk and bar light distributions are far less
peaked and generally do not dominate the center compared to the bulge,
therefore only the bulge model profile was corrected for seeing.

To account for the seeing effects, the model profiles of the bulge were
convolved with a Gaussian Point Spread Function (PSF) with the
same dispersion ($\sigma$) as the Gaussians that were fitted to some
field stars in the frame.  For a radially \mbox{symmetric} light distribution
around the center the seeing convolved profiles are described by
 \begin{equation}
\Sigma_{\rm s}(r) = \sigma^{-2} e^{-r^2/2\sigma^2} \int_0^\infty \Sigma(x) I_0(xr/\sigma^2) e^{x^2/2\sigma^2}x\,dx ,
\label{seecorr}
 \end{equation}
where $\Sigma(r)$ is the intrinsic surface brightness profile, $\sigma$
the dispersion of the Gaussian PSF and $I_0$ the zero-order modified
Bessel function of the first kind (Pritchet \& Kline~\cite{PriKli81}). 

The total model is simply the sum of the three components
 \begin{equation}
\Sigma_{\rm tot}(x,y) = \Sigma_{\rm disk}(x,y)+\Sigma_{\rm bulge}(x,y)+\Sigma_{\rm bar}(x,y).
 \end{equation}

\subsection{Fitting procedure}

To fit models to the data points a non-linear fitting algorithm capable
of accepting different weights for each data point was applied. 
Non-linear fitting algorithms are particularly sensitive to the initial
values provided, when searching for the minimum in the reduced $\chi^2$
of the fit. If the initial values are not ``reasonable'', the fitting
program can end up in a wrong local minimum. The results of the 1D
decompositions described in Sect.~\ref{compsect} were used as initial
values. The initial values for the bar were estimated by eye. The
routine generally converged to the same result to within the formal
errors, independent of the initial values. The formal fit errors were
usually much smaller than the errors due to the uncertainties in the
measurements and these errors will be discussed in detail in
Sect.~\ref{compsect} and~\ref{discus2}.

In decomposing 1D profiles it is common practice to fit in the
logarithmic (magnitude) regime. This means that one is effectively
trying to minimize the relative errors between model and data. 
Minimization in the logarithmic regime is not possible in the 2D case;
because of the noise some pixels will have negative values (below sky
level) in the outer parts of the galaxy image. To minimize relative
errors in the linear regime, a difference between model and data in a
low surface brightness region has to be given much more weight during
the fitting than the same difference in a high surface brightness area.
A weight function of the form ${\rm e}^{r/h}$ was used, where $h$ is
the initial estimate of the disk scalelength and $r$ is the
inclination corrected distance of the pixel from the center.   To
reduce computing time all pixels outside 2.5 initial disk scalelengths
were averaged over $5\times5$ pixels and given proportionally more
weight.

The 2D fitting procedure consisted of se\-ve\-ral steps.  First, all
images of a galaxy obtained in different passbands were aligned, freed
of foreground stars and their center was determined from the $R$
passband image (see Paper~I).  This center was fixed while fitting the
model components to the data.  After the fitting routine had converged,
the model light distribution was subtracted from the data.  The points
in the difference image that deviated more than 6 sigma (missed cosmic
ray events and faint stars) were flagged and not used in the next
iteration.  This process was repeated twice, each time taking the
initial estimates and the scalelength of the weight function from the
results of the previous step.  In general, the routine had already
converged to a satisfactory result after the second step. 

\subsection{Tests on artificial data}
\label{artim}

The 2D fitting routine was extensively tested on artificial images to
determine its reliability.  Artificial images are not really
representative of true galaxies, but can give an indication of the
systematic effects due to measurement errors.  By varying one by one the
observables in the artificial images one can investigate which measurement
error influences a model parameter the most.

The artificial images had the characteristics of a typical $R$~passband
observation:  all artificial galaxies had an exponential disk with a
\muo\ equal to 20 mag arcsec$^{-2}$, a scalelength of 20\arcsec\ and
axial ratio of 0.75. Just as with the typical observation the pixel
size was set at 0.3\arcsec, the seeing at 1.5\arcsec\ FWHM and the sky
surface brightness at 20 \magarc\ in the artificial images.  An
exponential bulge was added to each image, with the bulge parameters
chosen from a range in effective surface brightness and radius . The
images were created with photon and read out noise and a few areas were
set to undefined values to mimic the removal of foreground stars.  The
initial fit estimates for the disk and bulge parameters were set at
10-30\% off the intrinsic model values to check convergence.  The final
results showed very little sensitivity to the initial estimates. 

{
\tabcolsep=.8mm
\begin{table*}
{
\scriptsize
\caption[]
 {The relative errors between input artificial image parameters and
fitted bulge and disk parameters determined using the 2D fit method. 
Tabulated are the results for the standard artificial galaxy as
described in the text, as well as the effect of a wrong estimate of one
of the parameters which were normally kept fixed to the observed values
while fitting the model (see notes).  Relative errors are listed, with
$\Delta \mu_0 = \mu_{\rm 0,art}-\mu_{\rm 0,fit}$ and $\Delta \mu_{\rm e}
= \mu_{\rm e,art}-\mu_{\rm e,fit}$ in \magarc, $\Delta h = 2(h_{\rm
art}-h_{\rm fit})/(h_{\rm art}+h_{\rm fit})$ and $\Delta r_{\rm e} =
2(r_{\rm e,art}-r_{\rm e,fit})/(r_{\rm e,art}+r_{\rm e,fit})$
dimensionless. 
 }
\label{tst2d}

\begin{tabular}{clrrrrrrrrr@{\ \ \ \ \ }rrrrrrrrr}
\hline
\hline
$\mu_{\rm e,art}$&$r_{\rm e,art}$& \multicolumn{9}{c}{$\Delta \mu_{\rm 0}$} & \multicolumn{9}{c}{$\Delta h$}\\
   &     &stand.&seeing& \multicolumn{2}{c}{seeing error}&\multicolumn{2}{c}{sky error}&\multicolumn{2}{c}{$b/a$ error}&\multicolumn{1}{c}{PA\ \ }
         &stand.&seeing& \multicolumn{2}{c}{seeing error}&\multicolumn{2}{c}{sky error}&\multicolumn{2}{c}{$b/a$ error}&\multicolumn{1}{c}{PA}
\\
   &     &      &2.5\arcsec& --0.1\arcsec&+0.1\arcsec&+1\%&--1\%&--0.1&+0.1&+20\degr
         &      &2.5\arcsec& --0.1\arcsec&+0.1\arcsec&+1\%&--1\%&--0.1&+0.1&+20\degr
\\
\multicolumn{1}{c}{(1)}&
\multicolumn{1}{l}{(2)}&
\multicolumn{1}{c}{(3)}&
\multicolumn{1}{c}{(4)}&
\multicolumn{1}{c}{(5)}&
\multicolumn{1}{c}{(6)}&
\multicolumn{1}{c}{(7)}&
\multicolumn{1}{c}{(8)}&
\multicolumn{1}{c}{(9)}&
\multicolumn{1}{c}{(10)}&
\multicolumn{1}{c}{(11)\ \ \ }\\
\hline
17 & 4.0 & -0.012 & -0.042 & -0.034 & 0.011 &  0.088 &  -0.090 & 0.034 &  -0.056 & -0.051 &   -0.005 & -0.016 & -0.013 & 0.004 &  0.083 &  -0.083 & 0.094 &  -0.093 & -0.023\\
18 & 4.0 & -0.005 & -0.019 & -0.015 & 0.004 &  0.094 &  -0.084 & 0.041 &  -0.050 & -0.044 &   -0.001 & -0.007 & -0.005 & 0.002 &  0.086 &  -0.080 & 0.096 &  -0.091 & -0.021\\
19 & 4.0 & -0.002 & -0.009 & -0.006 & 0.001 &  0.096 &  -0.082 & 0.043 &  -0.048 & -0.042 &   -0.001 & -0.004 & -0.002 & 0.001 &  0.086 &  -0.079 & 0.097 &  -0.090 & -0.020\\
20 & 4.0 & -0.002 & -0.005 & -0.003 & 0.000 &  0.095 &  -0.082 & 0.043 &  -0.049 & -0.042 &   -0.000 & -0.002 & -0.001 & 0.000 &  0.086 &  -0.080 & 0.097 &  -0.090 & -0.020\\
21 & 4.0 & -0.001 & -0.003 & -0.002 & -0.001 & 0.091 &  -0.085 & 0.041 &  -0.052 & -0.043 &   -0.000 & -0.001 & -0.000 & -0.000 & 0.084 &  -0.081 & 0.096 &  -0.092 & -0.021\\
17 & 3.0 & -0.013 & -0.032 & -0.029 & 0.004 &  0.070 &  -0.078 & 0.021 &  -0.046 & -0.043 &   -0.005 & -0.013 & -0.011 & 0.002 &  0.077 &  -0.078 & 0.089 &  -0.089 & -0.020\\
18 & 3.0 & -0.005 & -0.015 & -0.013 & 0.001 &  0.077 &  -0.073 & 0.028 &  -0.040 & -0.036 &   -0.002 & -0.006 & -0.005 & 0.001 &  0.079 &  -0.076 & 0.091 &  -0.087 & -0.018\\
19 & 3.0 & -0.003 & -0.008 & -0.006 & -0.001 & 0.079 &  -0.071 & 0.029 &  -0.038 & -0.034 &   -0.001 & -0.003 & -0.002 & 0.000 &  0.080 &  -0.075 & 0.092 &  -0.086 & -0.017\\
20 & 3.0 & -0.002 & -0.004 & -0.003 & -0.001 & 0.079 &  -0.071 & 0.030 &  -0.038 & -0.033 &   -0.001 & -0.002 & -0.001 & -0.000 & 0.080 &  -0.075 & 0.092 &  -0.086 & -0.017\\
21 & 3.0 & -0.002 & -0.003 & -0.002 & -0.002 & 0.077 &  -0.073 & 0.030 &  -0.039 & -0.034 &   -0.000 & -0.001 & -0.001 & -0.000 & 0.079 &  -0.076 & 0.092 &  -0.086 & -0.017\\
17 & 2.0 & -0.013 & -0.006 & -0.023 & -0.001 & 0.058 &  -0.070 & 0.013 &  -0.039 & -0.037 &   -0.005 & -0.003 & -0.009 & -0.000 & 0.072 &  -0.075 & 0.086 &  -0.086 & -0.018\\
18 & 2.0 & -0.006 & -0.004 & -0.011 & -0.002 & 0.064 &  -0.064 & 0.019 &  -0.034 & -0.031 &   -0.002 & -0.001 & -0.004 & -0.000 & 0.074 &  -0.072 & 0.088 &  -0.084 & -0.016\\
19 & 2.0 & -0.004 & -0.003 & -0.005 & -0.002 & 0.066 &  -0.061 & 0.022 &  -0.032 & -0.029 &   -0.001 & -0.001 & -0.002 & -0.000 & 0.075 &  -0.071 & 0.089 &  -0.083 & -0.015\\
20 & 2.0 & -0.003 & -0.003 & -0.003 & -0.002 & 0.066 &  -0.061 & 0.022 &  -0.031 & -0.028 &   -0.001 & -0.001 & -0.001 & -0.000 & 0.075 &  -0.071 & 0.089 &  -0.083 & -0.015\\
21 & 2.0 & -0.002 & -0.003 & -0.003 & -0.002 & 0.065 &  -0.063 & 0.022 &  -0.031 & -0.028 &   -0.001 & -0.001 & -0.001 & -0.001 & 0.074 &  -0.072 & 0.089 &  -0.083 & -0.015\\
17 & 1.0 & 0.001 &  0.016 &  -0.008 & 0.009 &  0.063 &  -0.050 & 0.023 &  -0.024 & -0.021 &   0.001 &  0.007 &  -0.003 & 0.004 &  0.074 &  -0.066 & 0.089 &  -0.080 & -0.012\\
18 & 1.0 & -0.001 & 0.004 &  -0.005 & 0.002 &  0.061 &  -0.053 & 0.020 &  -0.026 & -0.023 &   -0.000 & 0.002 &  -0.002 & 0.001 &  0.073 &  -0.067 & 0.088 &  -0.081 & -0.013\\
19 & 1.0 & -0.002 & -0.001 & -0.004 & -0.001 & 0.059 &  -0.054 & 0.019 &  -0.027 & -0.024 &   -0.001 & -0.000 & -0.001 & -0.000 & 0.072 &  -0.068 & 0.088 &  -0.081 & -0.013\\
20 & 1.0 & -0.003 & -0.003 & -0.003 & -0.002 & 0.058 &  -0.055 & 0.018 &  -0.027 & -0.025 &   -0.001 & -0.001 & -0.001 & -0.001 & 0.072 &  -0.068 & 0.088 &  -0.081 & -0.013\\
21 & 1.0 & -0.003 & -0.004 & -0.003 & -0.003 & 0.057 &  -0.056 & 0.021 &  -0.028 & -0.025 &   -0.001 & -0.001 & -0.001 & -0.001 & 0.071 &  -0.069 & 0.089 &  -0.082 & -0.013\\
17 & 0.5 & 0.002 &  -0.001 & 0.000 &  0.003 &  0.061 &  -0.047 & 0.023 &  -0.021 & -0.018 &   0.001 &  -0.000 & 0.001 &  0.001 &  0.073 &  -0.065 & 0.089 &  -0.079 & -0.010\\
18 & 0.5 & -0.001 & -0.003 & -0.002 & -0.001 & 0.057 &  -0.050 & 0.019 &  -0.025 & -0.022 &   -0.000 & -0.001 & -0.000 & -0.000 & 0.071 &  -0.066 & 0.088 &  -0.080 & -0.012\\
19 & 0.5 & -0.003 & -0.004 & -0.003 & -0.003 & 0.056 &  -0.051 & 0.017 &  -0.026 & -0.023 &   -0.001 & -0.002 & -0.001 & -0.001 & 0.071 &  -0.067 & 0.087 &  -0.081 & -0.013\\
20 & 0.5 & -0.003 & -0.005 & -0.003 & -0.003 & 0.055 &  -0.053 & 0.017 &  -0.027 & -0.024 &   -0.001 & -0.002 & -0.001 & -0.001 & 0.071 &  -0.068 & 0.087 &  -0.081 & -0.013\\
21 & 0.5 & -0.003 & -0.005 & -0.003 & -0.003 & 0.055 &  -0.156 & 0.017 &  -0.027 & -0.264 &   -0.001 & -0.002 & -0.001 & -0.001 & 0.071 &  -0.104 & 0.087 &  -0.081 & 0.017 \\
\hline
\hline
\ \\
\hline
\hline
$\mu_{\rm e,art}$&$r_{\rm e,art}$& \multicolumn{9}{c}{$\Delta \mu_{\rm e}$} & \multicolumn{9}{c}{$\Delta r_{\rm e}$}\\
   &     &stand.& seeing& \multicolumn{2}{c}{seeing error}&\multicolumn{2}{c}{sky error}&\multicolumn{2}{c}{$b/a$ error}&\multicolumn{1}{c}{PA\ \ }
         &stand.& seeing& \multicolumn{2}{c}{seeing error}&\multicolumn{2}{c}{sky error}&\multicolumn{2}{c}{$b/a$ error}&\multicolumn{1}{c}{PA}
\\
   &     &      &2.5\arcsec& --0.1\arcsec&+0.1\arcsec&+1\%&--1\%&--0.1&+0.1&+20\degr
         &      &2.5\arcsec& --0.1\arcsec&+0.1\arcsec&+1\%&--1\%&--0.1&+0.1&+20\degr
\\
\hline
17 & 4.0 & -0.007 & -0.021 & -0.019 & 0.004 &  -0.007 & -0.008 & -0.006 & -0.009 & -0.008 &   -0.010 & -0.029 & -0.016 & -0.004 & -0.006 & -0.013 & -0.007 & -0.012 & -0.012\\
18 & 4.0 & -0.007 & -0.022 & -0.019 & 0.004 &  -0.005 & -0.009 & -0.004 & -0.010 & -0.010 &   -0.010 & -0.029 & -0.016 & -0.003 & -0.000 & -0.017 & -0.003 & -0.016 & -0.014\\
19 & 4.0 & -0.008 & -0.024 & -0.019 & 0.004 &  -0.002 & -0.012 & 0.001 &  -0.015 & -0.013 &   -0.009 & -0.029 & -0.015 & -0.003 & 0.014 &  -0.027 & 0.008 &  -0.025 & -0.021\\
20 & 4.0 & -0.008 & -0.029 & -0.019 & 0.003 &  0.005 &  -0.021 & 0.013 &  -0.029 & -0.023 &   -0.007 & -0.029 & -0.014 & -0.001 & 0.049 &  -0.054 & 0.033 &  -0.049 & -0.038\\
21 & 4.0 & -0.010 & -0.041 & -0.021 & 0.002 &  0.021 &  -0.046 & 0.037 &  -0.073 & -0.050 &   -0.004 & -0.029 & -0.011 & 0.002 &  0.132 &  -0.129 & 0.090 &  -0.121 & -0.087\\
17 & 3.0 & -0.011 & -0.046 & -0.028 & 0.006 &  -0.010 & -0.012 & -0.010 & -0.012 & -0.012 &   -0.017 & -0.050 & -0.026 & -0.009 & -0.014 & -0.020 & -0.015 & -0.019 & -0.019\\
18 & 3.0 & -0.011 & -0.046 & -0.028 & 0.007 &  -0.007 & -0.014 & -0.007 & -0.014 & -0.013 &   -0.017 & -0.050 & -0.026 & -0.008 & -0.007 & -0.025 & -0.011 & -0.022 & -0.021\\
19 & 3.0 & -0.010 & -0.048 & -0.027 & 0.007 &  0.000 &  -0.019 & -0.001 & -0.019 & -0.017 &   -0.016 & -0.050 & -0.024 & -0.007 & 0.010 &  -0.036 & -0.001 & -0.030 & -0.027\\
20 & 3.0 & -0.010 & -0.053 & -0.027 & 0.008 &  0.014 &  -0.033 & 0.012 &  -0.033 & -0.027 &   -0.014 & -0.048 & -0.022 & -0.005 & 0.048 &  -0.067 & 0.023 &  -0.051 & -0.043\\
21 & 3.0 & -0.010 & -0.068 & -0.028 & 0.008 &  0.045 &  -0.078 & 0.041 &  -0.078 & -0.057 &   -0.009 & -0.046 & -0.018 & 0.000 &  0.141 &  -0.155 & 0.077 &  -0.115 & -0.088\\
17 & 2.0 & -0.032 & -0.105 & -0.061 & -0.003 & -0.030 & -0.035 & -0.031 & -0.034 & -0.034 &   -0.039 & -0.087 & -0.054 & -0.026 & -0.034 & -0.044 & -0.038 & -0.042 & -0.042\\
18 & 2.0 & -0.031 & -0.105 & -0.060 & -0.001 & -0.024 & -0.037 & -0.027 & -0.035 & -0.035 &   -0.039 & -0.087 & -0.053 & -0.025 & -0.026 & -0.049 & -0.033 & -0.044 & -0.044\\
19 & 2.0 & -0.030 & -0.105 & -0.059 & 0.000 &  -0.012 & -0.045 & -0.020 & -0.040 & -0.038 &   -0.036 & -0.083 & -0.051 & -0.022 & -0.005 & -0.062 & -0.022 & -0.051 & -0.049\\
20 & 2.0 & -0.027 & -0.107 & -0.057 & 0.004 &  0.015 &  -0.067 & -0.002 & -0.054 & -0.048 &   -0.032 & -0.077 & -0.046 & -0.017 & 0.043 &  -0.099 & 0.004 &  -0.070 & -0.063\\
21 & 2.0 & -0.020 & -0.112 & -0.052 & 0.012 &  0.077 &  -0.141 & 0.037 &  -0.099 & -0.079 &   -0.021 & -0.062 & -0.038 & -0.005 & 0.157 &  -0.208 & 0.062 &  -0.127 & -0.104\\
17 & 1.0 & -0.121 & -0.115 & -0.209 & -0.022 & -0.112 & -0.130 & -0.118 & -0.125 & -0.125 &   -0.096 & -0.072 & -0.144 & -0.045 & -0.086 & -0.105 & -0.093 & -0.101 & -0.100\\
18 & 1.0 & -0.121 & -0.106 & -0.209 & -0.022 & -0.096 & -0.142 & -0.112 & -0.131 & -0.130 &   -0.096 & -0.064 & -0.142 & -0.043 & -0.070 & -0.118 & -0.086 & -0.105 & -0.104\\
19 & 1.0 & -0.120 & -0.086 & -0.209 & -0.021 & -0.060 & -0.172 & -0.097 & -0.146 & -0.142 &   -0.095 & -0.048 & -0.142 & -0.042 & -0.032 & -0.148 & -0.072 & -0.120 & -0.116\\
20 & 1.0 & -0.120 & -0.042 & -0.211 & -0.019 & 0.023 &  -0.250 & -0.064 & -0.184 & -0.172 &   -0.092 & -0.009 & -0.140 & -0.038 & 0.058 &  -0.227 & -0.035 & -0.157 & -0.144\\
21 & 1.0 & -0.142 & 0.050 &  -0.215 & -0.016 & 0.191 &  -0.501 & 1.107 &  -0.282 & -0.243 &   -0.101 & 0.078 &  -0.135 & -0.031 & 0.246 &  -0.469 & 0.717 &  -0.249 & -0.213\\
17 & 0.5 & 0.001 &  0.115 &  -0.368 & 0.467 &  0.052 &  -0.041 & 0.018 &  -0.018 & -0.014 &   -0.028 & 0.037 &  -0.214 & 0.198 &  0.010 &  -0.053 & -0.014 & -0.037 & -0.037\\
18 & 0.5 & -0.033 & 0.125 &  -0.390 & 0.417 &  0.087 &  -0.139 & 0.008 &  -0.082 & -0.073 &   -0.043 & 0.047 &  -0.227 & 0.174 &  0.037 &  -0.115 & -0.018 & -0.075 & -0.071\\
19 & 0.5 & -0.111 & 0.145 &  -0.441 & 0.305 &  0.169 &  -0.372 & -0.015 & -0.229 & -0.207 &   -0.084 & 0.073 &  -0.253 & 0.116 &  0.105 &  -0.262 & -0.018 & -0.163 & -0.150\\
20 & 0.5 & -0.270 & 0.175 &  -0.547 & 0.072 &  0.340 &  -0.894 & -0.192 & -0.536 & -0.478 &   -0.167 & 0.121 &  -0.308 & 0.000 &  0.257 &  -0.613 & -0.053 & -0.357 & -0.316\\
21 & 0.5 & -0.529 & 0.192 &  -0.750 & -0.304 & 1.016 &  -2.653 & -0.479 & -1.122 & -2.505 &   -0.301 & 0.227 &  -0.408 & -0.192 & 0.865 &  -1.851 & -0.167 & -0.726 & -1.936\\
\hline
\hline
\end{tabular}
}
{\em Notes:}\\
(1) The bulge effective surface brightness of the artificial galaxy in \magarc.\\
(2) The bulge effective radius of the artificial galaxy in arcsec.\\
(3) Fit results using the standard artificial image (see text) with $\muo=20$ $R$-\magarc\ and $h=20$\arcsec.\\
(4) As (3), but artificial image and fitting model have seeing of
2.5\arcsec\ instead of 1.5\arcsec.\\ 
(5) As (3), but fitting model has 1.4\arcsec\ seeing instead of the
1.5\arcsec\ of the artificial image.\\
(6) As (3), but fitting model has 1.6\arcsec\ seeing instead of
1.5\arcsec.\\
(7) As (3), but with sky level of the fitting model 1\% too low compared
to the real value in the artificial image. \\
(8) As (3), but with sky level of the fitting model 1\% too high compared
to the real value in the artificial image. \\
(9) As (3), but with $b/a=0.65$ in fitting model and 0.75 in the
artificial image.\\
(10) As (3), but with $b/a=0.85$ in fitting model and 0.75 in the
artificial image.\\
(11) As (3), with error of 20\degr\ in PA between image and model.
\end{table*}
}

\begin{table*}
\tabcolsep=1.35mm
\caption[]{
 The test results of the 2D fit using an $R$ passband image of UGC~438. 
Before the test a bulge with $\mu_{\rm e}$=18.954 $R$-\magarc\ and
$r_{\rm e}$=2.5\arcsec\ (the parameters resulting from the initial fit)
was subtracted from the original image, leaving a bulgeless test image. 
The removed bulge was for the test replaced by an exponential bulge of
the indicated model parameters (model \mue, model \re) and fitted by the
2D fit routine.  The resulting fitted $\mu_0$ and $\mu_{\rm e}$ are in
$R$-\magarc\ and the $h$ and $r_{\rm e}$ are in arcsec.  NC indicates
that no fit convergence was reached. 
 }
\label{testfitreal}
\begin{tabular}{c|@{\ \ \ \ \ \ \ }rccc@{\ \ \ \ \ \ \ \ \ \ \ \ \ }cccc@{\ \ \ \ \ \ \ \ \ \ \ \ \ }cccc}
\hline
\hline
model $\mu_{\rm e}$ & \multicolumn{4}{c}{model $r_{\rm e}$ = 2.5\arcsec\ \ \ \ \ \ }&\multicolumn{4}{c}{model $r_{\rm e}$ = 5.0\arcsec\ \ \ \ \ \ }&\multicolumn{4}{c}{model $r_{\rm e}$ = 1.25\arcsec}\\
        &fitted\ \ $\mu_0$\ \ \ & $h$ & $\mu_{\rm e}$ &  $r_{\rm e}$& $\mu_0$ & $h$ & $\mu_{\rm e}$ &  $r_{\rm e}$& $\mu_0$ & $h$ & $\mu_{\rm e}$ &  $r_{\rm e}$\\
\hline
 16.454 & 18.891 & 12.91 & 16.453 & 2.502 & 19.025 & 13.42 & 16.457 & 5.054 & 18.869 & 12.81 & 16.373 & 1.192\\
 17.207 & 18.891 & 12.91 & 17.205 & 2.502 & 19.026 & 13.42 & 17.213 & 5.087 & 18.869 & 12.81 & 17.037 & 1.125\\
 18.201 & 18.891 & 12.91 & 18.201 & 2.502 & 19.026 & 13.42 & 18.216 & 5.205 & 18.870 & 12.81 & 17.719 & 0.923\\
 18.954 & 18.891 & 12.91 & 18.954 & 2.502 & 19.025 & 13.42 & 18.977 & 5.406 & 18.873 & 12.83 & 17.792 & 0.638\\
 19.707 & 18.891 & 12.91 & 19.703 & 2.485 & 19.020 & 13.40 & 19.735 & 5.725 & 18.882 & 12.87 & 17.359 & 0.353\\
 20.459 & 18.888 & 12.90 & 20.415 & 2.401 & 18.997 & 13.31 & 20.454 & 6.011 & \multicolumn{4}{c}{ NC }\\
\hline
\hline
\end{tabular}

\end{table*}

The effect of different bulge-to-disk ratios was tested using
the artificial images with the different bulge parameter values. The
relative differences between artificial image input parameters and the
model fit output parameters are listed in Table~\ref{tst2d}.  The fitting
routine reproduced the bulge and disk parameters to a very high degree
of accuracy.  The disk parameters were reproduced with less than 1.5\%
error.  The bulge parameters were less well reproduced, especially for
very small effective radii and low effective surface brightnesses. 
These bulges are so small that they are unresolved with 1.5\arcsec\
seeing in the small region where they are dominating over the disk
surface brightness.  For these extreme cases it is better to convolve
not only the bulge model with the seeing, but also the disk. 
Increasing the seeing to 2.5\arcsec\ in both the artificial image and
the fitting model had no effect on the determination of the disk parameters
and slightly decreased the fit performance for the bulge parameters, as
can be seen in Table~\ref{tst2d}. 

Some observables measured from the images (seeing, sky background
level, $b/a$ and PA) were used as fixed parameters in the fit models. 
To estimate their relative effects on the derived parameters, all
artificial images were fitted again, but the fixed parameters in the
fitting model were given a wrong value with respect to the true values
in the artificial image. The typical maximum error was used for each of
the fixed parameters.  The test results can be found in
Table~\ref{tst2d}.  From these tests the largest error in \muo\ is
expected to result from a wrong estimation of the sky background.  The
error will be of order 0.1~\magarc, increasing, of course, for lower
surface brightness galaxies.  The errors in the scalelength are
dominated by errors in sky brightness and $b/a$, and both result in
errors less than 10\% in the typical case.  The errors in the bulge
parameters are dominated by their brightness and scalelength relative
to the disk parameters.  The parameters of the brighter bulges are most
effected by errors in the seeing.  The effective radii of the lower
surface brightness bulges are also influenced by sky brightness errors.
 Except for the extreme low surface brightness bulges and for bulges
with effective radii smaller than 1\arcsec\ the errors are never larger
than ~20\%. 


The fitting routine was also tested on some artificial images with a bar,
but with eight free parameters only a limited parameter space could be
explored. In general, the routine converged to satisfactory results in
just a few iterations.
  

\subsection{Tests on UGC\,438}

Clearly, artificial images are not a good representation of real
galaxies. Therefore, the fitting routine was also tested for different
bulge-to-disk ratios on an $R$ passband image of UGC\,438.  A 2D fitted
bulge was subtracted from the original image and replaced by artificial
exponential bulges with a range in effective surface brightness and
radius. The fit results for the different bulge parameters are listed
in Table~\ref{testfitreal}. 

The fitting routine generally reproduces the parameters quite well. 
The disk parameters change little for the different bulges (\muo\
changes at most by 0.04 \magarc\ and $h$ by about 4\%), but as with the
tests on artificial images, the fitting routine breaks down for bulges
with a small effective radius and a low surface brightness.  Next to
the explanations for this effect mentioned earlier for the artificial
images, one has another problem when one tests on real images. The
model bulge that was originally subtracted to create a bulgeless image
might not be perfectly correct. One is then left with some residuals
which have the greatest influence when later replaced by a low surface
brightness bulge.

\subsection{Resulting parameters}

{
\tabcolsep=1.55mm

\def\dsss{\multicolumn{1}{c}{---}}
\def\dssss{\multicolumn{1}{c}{---\ \ \ \ \ \ \ \ }}
\def\NP{\multicolumn{1}{c}{NP}}
\def\NC{\multicolumn{1}{c}{NC}}

\tablefirsthead{
\hline
\hline
\ \ \ UGC & \multicolumn{8}{c}{$B$\ \ } & \multicolumn{8}{c}{$K$}\\ 
 \phantom{12345} & \multicolumn{1}{c}{$\mu_0$} & \multicolumn{1}{c}{$h$} 
    & \multicolumn{1}{c}{$\mu_{\rm e}$} & \multicolumn{1}{c}{$r_{\rm e}$} 
    & \multicolumn{1}{c}{$\mu_{\rm 0,bar}$} & \multicolumn{1}{c}{$a_{\rm bar}$} 
    & \multicolumn{1}{c}{$b_{\rm bar}$} & \multicolumn{1}{l}{PA$_{\rm bar}$} 
    & \multicolumn{1}{c}{$\mu_0$} & \multicolumn{1}{c}{$h$} 
    & \multicolumn{1}{c}{$\mu_{\rm e}$} & \multicolumn{1}{c}{$r_{\rm e}$} 
    & \multicolumn{1}{c}{$\mu_{\rm 0,bar}$} & \multicolumn{1}{c}{$a_{\rm bar}$} 
    & \multicolumn{1}{c}{$b_{\rm bar}$} & \multicolumn{1}{c}{PA$_{\rm bar}$} \\
\hline
}

\tablehead{
\multicolumn{17}{l}{\small{\bf \hspace{-1.7mm}Table~\ref{bd4fpar}.}~-continued.}\\
\multicolumn{17}{l}{\ }\\
\hline
\hline
\ \ \ UGC & \multicolumn{8}{c}{$B$\ \ } & \multicolumn{8}{c}{$K$}\\ 
    & \multicolumn{1}{c}{$\mu_0$} & \multicolumn{1}{c}{$h$} 
    & \multicolumn{1}{c}{$\mu_{\rm e}$} & \multicolumn{1}{c}{$r_{\rm e}$} 
    & \multicolumn{1}{c}{$\mu_{\rm 0,bar}$} & \multicolumn{1}{c}{$a_{\rm bar}$} 
    & \multicolumn{1}{c}{$b_{\rm bar}$} & \multicolumn{1}{l}{PA$_{\rm bar}$} 
    & \multicolumn{1}{c}{$\mu_0$} & \multicolumn{1}{c}{$h$} 
    & \multicolumn{1}{c}{$\mu_{\rm e}$} & \multicolumn{1}{c}{$r_{\rm e}$} 
    & \multicolumn{1}{c}{$\mu_{\rm 0,bar}$} & \multicolumn{1}{c}{$a_{\rm bar}$} 
    & \multicolumn{1}{c}{$b_{\rm bar}$} & \multicolumn{1}{c}{PA$_{\rm bar}$} \\
\hline
}

\tabletail{
\hline \hline
}

\def\cap{
 The results from the 2D fit model with $\mu_0$, $\mu_{\rm e}$ and
$\mu_{\rm 0, bar}$ in \magarc, $h$, $r_{\rm e}$, $a_{\rm bar}$ and
$b_{\rm bar}$ in arcsec and PA$_{\rm bar}$ in degrees.  NP
means a non-photometric observation, but the scale
parameters are still determined. 
 }
\tablecaption{
\cap
\label{bd4fpar}
}

\begin{supertabularts}{r@{\ \ \ \ \ \ }rrrrrrrr@{\ \ \ \ \ \ \ \ \ }rrrrrrrr}{\rm}
      89 & 22.07  & 28.5  & 18.87  & 2.5   & 21.77  & 31.6   & 11.4   & 152.8  & 17.46  & 17.4  & 14.61  & 2.5   & 17.63  & 29.6   & 10.3   & 153.6  \\
      93 & 22.33  & 21.3  & 22.02  & 0.9   & \dsss  & \dsss  & \dsss  & \dssss & 18.55  & 15.7  & 18.82  & 1.4   & \dsss  & \dsss  & \dsss  & \dsss  \\
     242 & 21.26  & 13.7  & 18.73  & 0.2   & \dsss  & \dsss  & \dsss  & \dssss & 17.28  & 11.1  & 17.82  & 0.9   & \dsss  & \dsss  & \dsss  & \dsss  \\
     334 & 23.36  & 21.5  & 24.89  & 2.4   & \dsss  & \dsss  & \dsss  & \dssss & 20.32  & 18.1  & 21.26  & 3.8   & \dsss  & \dsss  & \dsss  & \dsss  \\
     438 & 20.45  & 14.1  & 21.22  & 2.9   & \dsss  & \dsss  & \dsss  & \dssss & 16.23  & 11.5  & 15.84  & 2.1   & \dsss  & \dsss  & \dsss  & \dsss  \\
     463 & 20.76  & 13.5  & 20.60  & 1.3   & \dsss  & \dsss  & \dsss  & \dssss & 16.80  & 12.0  & 16.73  & 1.9   & \dsss  & \dsss  & \dsss  & \dsss  \\
     490 & 21.47  & 18.5  & 21.51  & 2.6   & \dsss  & \dsss  & \dsss  & \dssss & 17.16  & 14.9  & 17.73  & 3.9   & \dsss  & \dsss  & \dsss  & \dsss  \\
     508 & 22.05  & 30.5  & 19.81  & 2.8   & 22.66  & 48.1   & 11.2   & 93.1   & 17.74  & 26.3  & 15.26  & 2.9   & 17.91  & 40.1   & 10.5   & 93.3   \\
     628 & 22.86  & 14.8  & 24.61  & 3.4   & \dsss  & \dsss  & \dsss  & \dssss & 20.39  & 18.6  & 21.46  & 5.4   & \dsss  & \dsss  & \dsss  & \dsss  \\
    1305 & 22.02  & 33.9  & 21.24  & 3.8   & \dsss  & \dsss  & \dsss  & \dssss & 17.61  & 26.5  & 17.00  & 4.2   & \dsss  & \dsss  & \dsss  & \dsss  \\
    1455 & 22.27  & 24.0  & 20.47  & 2.1   & 22.15  & 13.2   & 4.9    & 19.6   & 17.64  & 16.2  & 15.67  & 2.0   & 18.06  & 16.5   & 6.2    & 22.5   \\
    1551 & 22.47  & 25.8  & 24.81  & 2.6   & \dsss  & \dsss  & \dsss  & \dssss & 18.98  & 24.4  & 20.06  & 2.9   & \dsss  & \dsss  & \dsss  & \dsss  \\
    1559 & 22.48  & 20.0  & 23.42  & 3.8   & \dsss  & \dsss  & \dsss  & \dssss & 20.08  & 19.6  & 20.21  & 5.8   & \dsss  & \dsss  & \dsss  & \dsss  \\
    1577 & 22.44  & 22.1  & 20.89  & 2.2   & 22.80  & 32.9   & 10.8   & 144.5  & 18.26  & 17.6  & 16.01  & 1.8   & 18.42  & 26.8   & 8.9    & 145.1  \\
    1719 & 22.45  & 21.7  & 20.99  & 1.6   & 22.88  & 13.3   & 5.2    & 127.6  & 17.73  & 14.3  & 16.68  & 1.9   & 18.45  & 15.9   & 5.2    & 129.3  \\
    1792 & 21.65  & 17.1  & 21.31  & 1.3   & \dsss  & \dsss  & \dsss  & \dssss & 17.48  & 14.4  & 17.21  & 1.8   & \dsss  & \dsss  & \dsss  & \dsss  \\
    2064 & 22.28  & 20.1  & 20.78  & 0.8   & \dsss  & \dsss  & \dsss  & \dssss & 18.01  & 17.6  & 16.25  & 1.1   & \dsss  & \dsss  & \dsss  & \dsss  \\
    2081 & 22.31  & 19.1  & 23.46  & 1.4   & \dsss  & \dsss  & \dsss  & \dssss & 19.44  & 19.4  & 19.83  & 2.5   & \dsss  & \dsss  & \dsss  & \dsss  \\
    2124 & 22.34  & 23.0  & 20.07  & 2.7   & 21.96  & 21.4   & 7.0    & 100.1  & 18.11  & 21.3  & 15.89  & 2.7   & 17.62  & 21.3   & 6.1    & 100.2  \\
    2125 & 23.20  & 26.8  & 21.01  & 1.7   & 23.47  & 34.6   & 6.9    & 79.5   & \dsss  & \dsss & \dsss  & \dsss & \dsss  & \dsss  & \dsss  & \dsss  \\
    2197 & 22.57  & 18.2  & 23.43  & 2.1   & \dsss  & \dsss  & \dsss  & \dssss & 17.97  & 12.3  & 18.60  & 2.1   & \dsss  & \dsss  & \dsss  & \dsss  \\
    2368 & 23.67  & 28.3  & 19.82  & 1.3   & 22.50  & 52.9   & 14.0   & 156.9  & 18.88  & 18.1  & 15.19  & 1.7   & 18.41  & 48.0   & 11.9   & 156.9  \\
    2595 & \dsss  & \dsss & \dsss  & \dsss & \dsss  & \dsss  & \dsss  & \dssss & \dsss  & \dsss & \dsss  & \dsss & \dsss  & \dsss  & \dsss  & \dsss  \\
    3066 & 22.03  & 14.6  & 21.40  & 0.4   & \dsss  & \dsss  & \dsss  & \dssss & 17.09  & 11.2  & 17.25  & 1.6   & \dsss  & \dsss  & \dsss  & \dsss  \\
    3080 & 21.99  & 17.2  & 19.88  & 0.2   & \dsss  & \dsss  & \dsss  & \dssss & 18.21  & 15.1  & 18.79  & 1.7   & \dsss  & \dsss  & \dsss  & \dsss  \\
    3140 & 20.90  & 13.1  & 20.46  & 2.0   & \dsss  & \dsss  & \dsss  & \dssss & 16.96  & 11.3  & 15.99  & 2.1   & \dsss  & \dsss  & \dsss  & \dsss  \\
    4126 & 21.83  & 22.6  & 20.01  & 1.4   & 23.34  & 45.7   & 9.6    & 38.1   & 18.02  & 18.7  & 16.09  & 1.8   & 19.01  & 35.2   & 10.8   & 48.5   \\
    4256 & 21.18  & 17.6  & 19.22  & 0.7   & \dsss  & \dsss  & \dsss  & \dssss & 17.27  & 15.8  & 16.15  & 2.1   & \dsss  & \dsss  & \dsss  & \dsss  \\
    4308 & 21.34  & 20.0  & 19.78  & 0.8   & \dsss  & \dsss  & \dsss  & \dssss & 17.66  & 16.4  & 17.06  & 1.9   & \dsss  & \dsss  & \dsss  & \dsss  \\
    4368 & 21.52  & 17.6  & 22.52  & 2.0   & \dsss  & \dsss  & \dsss  & \dssss & 17.81  & 14.7  & 18.89  & 2.2   & \dsss  & \dsss  & \dsss  & \dsss  \\
    4375 & 21.31  & 20.2  & 22.18  & 2.1   & \dsss  & \dsss  & \dsss  & \dssss & 17.21  & 16.9  & 18.46  & 2.5   & \dsss  & \dsss  & \dsss  & \dsss  \\
    4422 & 22.04  & 31.0  & 20.92  & 3.5   & \dsss  & \dsss  & \dsss  & \dssss & 18.38  & 27.0  & 16.55  & 3.4   & \dsss  & \dsss  & \dsss  & \dsss  \\
    4458 & 21.72  & 19.8  & 19.50  & 3.0   & \dsss  & \dsss  & \dsss  & \dssss & 17.75  & 17.9  & 15.05  & 2.5   & \dsss  & \dsss  & \dsss  & \dsss  \\
    5103 & 20.50  & 17.8  & 19.84  & 1.3   & \dsss  & \dsss  & \dsss  & \dssss & 16.56  & 14.5  & 16.07  & 1.9   & \dsss  & \dsss  & \dsss  & \dsss  \\
    5303 & 21.32  & 36.8  & 21.41  & 3.0   & \dsss  & \dsss  & \dsss  & \dssss & 17.89  & 37.7  & 17.72  & 4.3   & \dsss  & \dsss  & \dsss  & \dsss  \\
    5510 & 20.66  & 20.4  & 19.22  & 1.3   & \dsss  & \dsss  & \dsss  & \dssss & 17.37  & 18.4  & 16.53  & 2.2   & \dsss  & \dsss  & \dsss  & \dsss  \\
    5554 & 20.98  & 19.1  & 19.31  & 1.5   & \dsss  & \dsss  & \dsss  & \dssss & 16.94  & 16.7  & 15.66  & 2.1   & \dsss  & \dsss  & \dsss  & \dsss  \\
    5633 & 23.13  & 27.8  & 23.67  & 0.9   & 22.99  & 23.5   & 3.0    & 161.5  & 20.01  & 29.4  & 21.03  & 1.5  & 19.88  & 23.2   & 3.7    & 162.5  \\
    5842 & 21.44  & 30.3  & 21.01  & 0.9   & \dsss  & \dsss  & \dsss  & \dssss & 18.01  & 27.4  & 18.52  & 2.2   & \dsss  & \dsss  & \dsss  & \dsss  \\
    6028 & 20.49  & 15.3  & 21.57  & 0.5   & \dsss  & \dsss  & \dsss  & \dssss & 17.27  & 13.0  & 18.66  & 1.6   & \dsss  & \dsss  & \dsss  & \dsss  \\
    6077 & 20.88  & 18.4  & 20.17  & 1.1   & \dsss  & \dsss  & \dsss  & \dssss & 17.49  & 18.4  & 17.22  & 2.1   & \dsss  & \dsss  & \dsss  & \dsss  \\
    6123 & \NP    & 27.1  & \NP    & 2.1   & \dsss  & \dsss  & \dsss  & \dssss & 17.29  & 25.0  & 15.46  & 2.0   & \dsss  & \dsss  & \dsss  & \dsss  \\
    6277 & 20.88  & 27.9  & 18.98  & 1.3   & \dsss  & \dsss  & \dsss  & \dssss & 17.25  & 21.5  & 16.54  & 2.9   & \dsss  & \dsss  & \dsss  & \dsss  \\
    6445 & 20.55  & 16.1  & 20.19  & 2.6   & 20.94  & 11.6   & 4.1    & 157.6  & 16.83  & 14.0  & 16.17  & 2.6   & 17.11  & 11.4   & 4.5    & 154.4  \\
    6453 & 20.91  & 21.3  & 22.04  & 7.9   & \dsss  & \dsss  & \dsss  & \dssss & 17.16  & 16.3  & 17.87  & 5.6   & \dsss  & \dsss  & \dsss  & \dsss  \\
    6460 & 20.66  & 27.6  & 19.63  & 1.8   & \dsss  & \dsss  & \dsss  & \dssss & 17.15  & 24.1  & 16.53  & 2.1   & \dsss  & \dsss  & \dsss  & \dsss  \\
    6536 & \NP    & 21.7  & \NP    & 1.9   & \NP    & 13.5   & 9.3    & 9.0    & 18.57  & 20.3  & 15.47  & 1.9   & 17.80  & 12.7   & 8.1    & 16.5   \\
    6693 & 21.60  & 18.9  & 21.72  & 0.7   & \dsss  & \dsss  & \dsss  & \dssss & 17.99  & 16.7  & 17.13  & 1.3   & \dsss  & \dsss  & \dsss  & \dsss  \\
    6746 & 21.38  & 17.7  & 20.18  & 2.2   & \dsss  & \dsss  & \dsss  & \dssss & 17.30  & 16.0  & 15.64  & 2.1   & \dsss  & \dsss  & \dsss  & \dsss  \\
    6754 & 22.71  & 32.0  & 21.58  & 4.2   & \dsss  & \dsss  & \dsss  & \dssss & 18.32  & 21.7  & 15.77  & 1.8   & \dsss  & \dsss  & \dsss  & \dsss  \\
    7169 & 20.11  & 13.2  & 20.20  & 1.7   & \dsss  & \dsss  & \dsss  & \dssss & 16.55  & 10.0  & 16.38  & 2.3   & \dsss  & \dsss  & \dsss  & \dsss  \\
    7315 & 19.99  & 14.8  & 20.75  & 1.5   & \dsss  & \dsss  & \dsss  & \dssss & 16.03  & 14.6  & 16.96  & 2.1   & \dsss  & \dsss  & \dsss  & \dsss  \\
    7450 & 21.17  & 64.9  & 20.14  & 7.8   & \dsss  & \dsss  & \dsss  & \dssss & 17.30  & 48.5  & 15.72  & 6.3   & \dsss  & \dsss  & \dsss  & \dsss  \\
    7523 & 21.49  & 36.0  & 19.61  & 4.0   & 22.10  & 47.2   & 11.8   & 144.5  & 17.70  & 32.5  & 15.11  & 3.1   & 17.87  & 44.1   & 12.9   & 143.6  \\
    7594 & 21.03  & 52.4  & 19.16  & 3.6   & 21.72  & 50.2   & 21.5   & 11.8   & 16.94  & 44.8  & 15.21  & 4.1   & 17.73  & 51.2   & 19.6   & 8.7    \\
    7876 & \NP    & 21.2  & \NP    & 1.7   & \dsss  & \dsss  & \dsss  & \dssss & 18.55  & 22.1  & 20.06  & 4.4   & \dsss  & \dsss  & \dsss  & \dsss  \\
    7901 & 20.07  & 25.1  & 20.06  & 3.0   & \dsss  & \dsss  & \dsss  & \dssss & 16.06  & 19.7  & 16.22  & 3.8   & \dsss  & \dsss  & \dsss  & \dsss  \\
    8279 & 20.52  & 13.6  & 22.87  & 7.0   & \dsss  & \dsss  & \dsss  & \dssss & 16.90  & 11.5  & 18.59  & 5.9   & \dsss  & \dsss  & \dsss  & \dsss  \\
    8289 & 21.80  & 33.2  & 19.59  & 3.2   & \dsss  & \dsss  & \dsss  & \dssss & 17.90  & 21.1  & 15.84  & 3.2   & \dsss  & \dsss  & \dsss  & \dsss  \\
    8865 & 21.89  & 32.0  & 20.10  & 2.7   & 21.83  & 19.8   & 9.4    & 169.2  & 18.27  & 29.3  & 15.65  & 2.3   & 17.84  & 27.2   & 9.5    & 168.4  \\
    9024 & 24.08  & 28.2  & 21.64  & 2.1   & \dsss  & \dsss  & \dsss  & \dssss & 22.07  & 45.4  & 19.15  & 3.6   & \dsss  & \dsss  & \dsss  & \dsss  \\
    9061 & 22.63  & 62.0  & 20.08  & 2.1   & 21.97  & 15.0   & 6.5    & 111.3  & 18.75  & 39.3  & 16.22  & 2.8   & 17.73  & 16.8   & 4.0    & 110.2  \\
    9481 & 21.22  & 17.6  & 21.36  & 1.1   & 22.87  & 27.0   & 4.3    & 94.2   & 18.03  & 17.5  & 17.26  & 1.6   & 18.46  & 21.5   & 6.2    & 98.7   \\
    9915 & \NP    & 17.7  & \NP    & 2.6   & \dsss  & \dsss  & \dsss  & \dssss & 17.21  & 15.7  & 16.58  & 2.3   & \dsss  & \dsss  & \dsss  & \dsss  \\
    9926 & 20.13  & 17.9  & 20.16  & 3.1   & \dsss  & \dsss  & \dsss  & \dssss & 16.45  & 16.6  & 15.96  & 4.3   & \dsss  & \dsss  & \dsss  & \dsss  \\
    9943 & 20.40  & 20.0  & 20.92  & 2.7   & 21.37  & 19.8   & 6.2    & 73.3   & 16.60  & 17.6  & 16.55  & 2.3   & 17.16  & 19.7   & 5.9    & 70.3   \\
   10083 & 21.51  & 23.3  & 22.65  & 4.3   & 22.08  & 44.8   & 8.8    & 149.2  & 17.60  & 18.9  & 18.64  & 4.2   & 18.50  & 44.2   & 10.1   & 149.3  \\
   10437 & 24.03  & 33.6  & 23.28  & 8.9   & \dsss  & \dsss  & \dsss  & \dssss & \NP    & 17.0  & \NP    & 7.1   & \dsss  & \dsss  & \dsss  & \dsss  \\
   10445 & 21.76  & 20.3  & 22.99  & 2.3   & \dsss  & \dsss  & \dsss  & \dssss & 19.00  & 19.6  & 20.08  & 4.5   & \dsss  & \dsss  & \dsss  & \dsss  \\
   10584 & 21.77  & 23.8  & 21.22  & 1.7   & \dsss  & \dsss  & \dsss  & \dssss & 18.18  & 19.4  & 17.47  & 2.2   & \dsss  & \dsss  & \dsss  & \dsss  \\
   11628 & 22.27  & 38.8  & 20.10  & 3.1   & \dsss  & \dsss  & \dsss  & \dssss & 17.19  & 21.9  & 15.09  & 2.6   & \dsss  & \dsss  & \dsss  & \dsss  \\
   11708 & 21.51  & 17.9  & 21.77  & 1.9   & \dsss  & \dsss  & \dsss  & \dssss & \NP    & 14.4  & \NP    & 2.0   & \dsss  & \dsss  & \dsss  & \dsss  \\
   11872 & 20.46  & 18.3  & 20.68  & 7.0   & 20.61  & 13.2   & 3.7    & 17.6   & 15.71  & 12.6  & 14.94  & 2.5   & 16.01  & 12.8   & 6.0    & 13.3   \\
   12151 & 23.27  & 26.1  & 24.78  & 6.4   & \dsss  & \dsss  & \dsss  & \dssss & 20.14  & 21.7  & 21.19  & 4.8   & \dsss  & \dsss  & \dsss  & \dsss  \\
   12343 & 21.95  & 50.7  & 21.32  & 4.3   & 22.01  & 66.9   & 11.0   & 7.0    & 17.67  & 39.7  & 15.94  & 3.4   & 17.67  & 68.2   & 12.6   & 10.7   \\
   12379 & 21.99  & 19.3  & 20.40  & 2.1   & \dsss  & \dsss  & \dsss  & \dssss & 17.44  & 15.7  & 15.33  & 1.9   & \dsss  & \dsss  & \dsss  & \dsss  \\
   12391 & 21.50  & 15.7  & 18.03  & 0.2   & \dsss  & \dsss  & \dsss  & \dssss & 17.79  & 13.8  & 17.36  & 1.2   & \dsss  & \dsss  & \dsss  & \dsss  \\
   12511 & 22.46  & 24.4  & 22.38  & 1.6   & \dsss  & \dsss  & \dsss  & \dssss & 18.35  & 13.2  & 18.09  & 2.1   & \dsss  & \dsss  & \dsss  & \dsss  \\
   12614 & 20.94  & 21.6  & 18.76  & 0.7   & \dsss  & \dsss  & \dsss  & \dssss & 17.24  & 20.2  & 15.00  & 1.2   & \dsss  & \dsss  & \dsss  & \dsss  \\
   12638 & 22.17  & 21.4  & 21.74  & 1.3   & \dsss  & \dsss  & \dsss  & \dssss & 18.27  & 19.8  & 18.24  & 2.6   & \dsss  & \dsss  & \dsss  & \dsss  \\
   12654 & 21.76  & 19.6  & 22.73  & 2.0   & \dsss  & \dsss  & \dsss  & \dssss & 17.98  & 15.2  & 18.82  & 3.0   & \dsss  & \dsss  & \dsss  & \dsss  \\
   12732 & 23.66  & 29.8  & 25.30  & 11.0  & \dsss  & \dsss  & \dsss  & \dssss & \NP    & 37.5  & \NP    & 10.5  & \dsss  & \dsss  & \dsss  & \dsss  \\
   12754 & 21.81  & 52.7  & 22.21  & 5.2   & 21.83  & 48.1   & 6.7    & 98.0   & \NP    & 49.2  & \NP    & 7.6   & \NP    & 44.6   & 8.3    & 100.6  \\
   12776 & 23.39  & 61.3  & 20.01  & 2.0   & 22.02  & 22.6   & 8.9    & 171.5  & 19.30  & 30.2  & 15.52  & 2.0   & 17.90  & 21.0   & 7.9    & 171.6  \\
   12808 & 20.31  & 13.8  & 17.50  & 0.7   & \dsss  & \dsss  & \dsss  & \dssss & \dsss  & \dsss & \dsss  & \dsss & \dsss  & \dsss  & \dsss  & \dsss  \\
   12845 & 22.69  & 24.6  & 23.32  & 2.4   & \dsss  & \dsss  & \dsss  & \dssss & \NP    & 18.1  & \NP    & 2.5   & \dsss  & \dsss  & \dsss  & \dsss  \\
\end{supertabularts}
}

\begin{figure*}
\vspace{11.7cm}
 \caption[]{
 Left: UGC~89 $R$ passband image, with isophotes overlaid from 18 to 24
$R$-\magarc\ in steps of 0.5 \magarc.  Center: Fitted model on the same
grayscale and contour level as the left image. Right: Residual image using the
2D fit model of the center image.  The ellipse indicates the 
area used for the fit and the $b/a$ and PA of the fitted
exponential disk.  The fitted bar component shows up as the inner
ellipse in the residuals.  Structure in the bar region and the two arms
coming off the ends of the bar are clearly visible and can not be fitted with
this simple model. 
}
 \label{u89fit}
 \end{figure*}

The results from the 2D fits for all galaxies in our sample are
presented in Table~\ref{bd4fpar} for the $B$ and the $K$ passband. This
table contains observed data values. No corrections were applied for
inclination and for internal and Galactic extinction. In 23 out of 86
cases it was found that the fits of at least the bulge improved if a
bar was also included.

Figure~\ref{u89fit} shows an example of an original, a model and a model
subtracted image of UGC\,89.  The relative differences between model and
data are quite small and the resulting difference image can be used to
study small scale structures in the bulge and bar region and to study
spiral arms.  UGC\,89 is a good example of a galaxy that can only be
fitted correctly by a 2D method with a bar, as can be seen more clearly
in the resulting 1D profile of Fig.~\ref{plfit}.  

In testing both artificial and real images it appears that the 2D
fitting routine has difficulties fitting bulges with a low surface
brightness and a small effective radius compared to the disk parameters. 
There are few galaxies in the sample with a fitted $r_{\rm e} \!< \!1$. 
This is only found in the $B$ and $V$ passbands when galaxies have
Type~II profiles.  The routine fits a small bulge to these galaxy images
to avoid filling up the ``central hole'' in the disk.  The Type~II
behavior is always less pronounced at longer wavelengths and indeed in
the $K$ passband there are no galaxies with $r_{\rm e} \!< \!1$. 
Fortunately, there are no galaxies in the sample with a small effective
radius {\em and} a low effective surface brightness of the bulge
relative to the disk central surface brightness.  Using these
limitations in parameter space, the standard column in Table~\ref{tst2d}
indicates that the errors intrinsic to the fitting routine are expected
to be less than 10\%. The errors caused by other sources will be
discussed in the next sections.

\section{Comparison of different decomposition methods}
\label{compsect}
 The structural parameters derived in the previous section will be used
in subsequent papers of this series. In order to assess the reliability
of the derived relations in these papers, a good estimate of systematic
and random errors in the fitted parameters has to be determined. The 2D
fitting results will be compared to various conventional 1D methods to
demonstrate the increased accuracy. The 1D methods will also be used to
investigate the two most important sources of error in the bulge and
disk parameters, namely the sky background error and the uncertainty in
the shape of the bulge profile. The 2D method was too time consuming to
be used for these tests.

\subsection{One-dimensional decompositions}

One-dimensional decomposition methods are well known in the literature
(Kormendy~\cite{Kor77}; Schombert \& Bothun~\cite{SchBot87};
Simien~\cite{Sim89}; Capaccioli \& Caon~\cite{CapCao92}; Andredakis \&
Sanders~\cite{AndSan94}) and will be described briefly.  The
resulting parameters are only available in electronic form. 

 The most elementary way to obtain the disk parameters, the ``marking
the disk method'', was also used by Freeman (\cite{Freeman}), when he
found the disk central surface brightness to be constant among
galaxies.  The linear part of the luminosity profile, plotted on a
magnitude scale, was marked and a linear least squares fit was made to
the data points in the indicated range. To be able to compare the disk
parameters of a galaxy in different passbands, I used the same range in
radii for all passbands.  The resulting parameters can be quite
sensitive to the minimum and maximum radius chosen to fit.  The
difficulties of the ``marking the disk'' method have been discussed by
Giovanelli et al.~(\cite{Gio94}) and results obtained by different
authors were compared by Knapen \& van der Kruit (\cite{KnavdK}),
showing remarkable differences. These differences were mainly caused by
a change in scalelength at a lower surface brightness.

The ``marking the disk'' method yields disk parameters that are
intuitively correct for the human eye. However, the luminosity profile
is a combination of bulge and disk light and to get correct results
both should be fitted simultaneously (Kormendy~\cite{Kor77}). The
numerical method to decompose the luminosity profiles used here is
essentially identical to the method described by Andredakis \& Sanders
(\cite{AndSan94}). A non-linear $\chi^2$ minimalization routine was used
to fit the model profiles to the data points in the logarithmic regime.
Both bulge and disk model profiles were convolved with the seeing PSF
using Eq.~(\ref{seecorr}) and the fits were limited by the same maximum
radii used for the ``marking the disk'' fit.

As already indicated in Sect.~\ref{Intro}, several different radial
luminosity laws have been proposed for the light of bulges.  One of the
more general forms for the light profile of the bulge is 
the generalized exponential law, originally proposed by S\'ersic
(\cite{Ser68}):
 \begin{equation}
\Sigma(r) = \Sigma_0{\rm e}^{-(r/h)^{1/n}}.
\label{genexp}
 \end{equation}
 This generalized exponential law has been applied to fit
elliptical and S0 galaxies by Caon et al.~(\cite{Caon93}) and to dwarf
ellipticals by Young \& Currie (\cite{YouCur94}). 

As a first step, the fitting of an exponential disk and a generalized
exponential bulge profile was tried. A wide variety of initial values
for the five free parameters were tried, but for most of the galaxies
the fits did not converge to physically acceptable values (negative
$\Sigma_0$, $\Sigma_{\rm e}$, $h$ or \re).  The value of $n$ in
Eq.~(\ref{genexp}) was of order 0.5--5 in the cases where the fit did
converge, that is for the galaxies with pronounced bulges. For most
galaxies the bulge light dominates over the disk light at only a few
data points and these few points do not carry enough information to
limit the shape parameter $n$.

To reach more stable results the same fits were made again with $n$
fixed to values 1, 2 and 4.  With $n \!= \!4$ and after redefining $\Sigma_0$
and $h$ into effective parameters, Eq.~(\ref{genexp}) translates in the
most commonly used bulge fitting function, the de Vaucouleurs
(\cite{deV48}) or \rq law:
 \begin{equation}
\Sigma(r) = \Sigma_{\rm e} {\rm e}^{-7.67({r/r_{\rm e}}^{1/4}-1)} .
\label{roneq}
 \end{equation}
Setting $n=2$ in Eq.~(\ref{genexp}) gives an ``\rh law'' profile:
 \begin{equation}
\Sigma(r) = \Sigma_{\rm e} {\rm e}^{-3.672({r/r_{\rm e}}^{1/2}-1)} .
 \end{equation}
 In the case of $n \!= \!1$ in Eq.~(\ref{genexp}) one has the exponential law
normally used for disk profiles, which translates into
Eq.~(\ref{expbul}) when rewritten to effective parameters and which was
also used for the 2D fit.

More stable results were reached than with the general exponential law,
but the intrinsic properties of the $r^{1/4}$ law still made it
impossible to reach convergence in many cases.  With $n \!= \!2$ the
situation improved considerably and using an exponential profile for
the bulge, the fitting converged for all galaxies except for UGC\,6028. 
This galaxy has a very small bulge in a Type~II profile (Freeman
\cite{Freeman}) and therefore the fitting routine tends to make the bulge
negative in order to create a hole in the disk profile.  The fits with
the exponential bulge are the most stable; convergence is reached more
often and the fit results are less sensitive to initial values with the
exponential bulge than with the other tested profiles. 

The tests on the artificial images in Sect.~\ref{artim} showed that
one of the main sources of error in the bulge and disk parameters was
the uncertainty in the sky background level.  The maximum errors in the
fit parameters due to this uncertainty were calculated for each galaxy
using the sky errors determined in Paper~I.  All 1D fits were repeated
with the maximum sky error estimate added to and subtracted from the
luminosity profile.  These error estimates will be used in the next
sections. 

 \subsection{Profile comparison}

\begin{figure*}
\mbox{\epsfxsize=16.8cm \epsfbox[45 138 530 763]{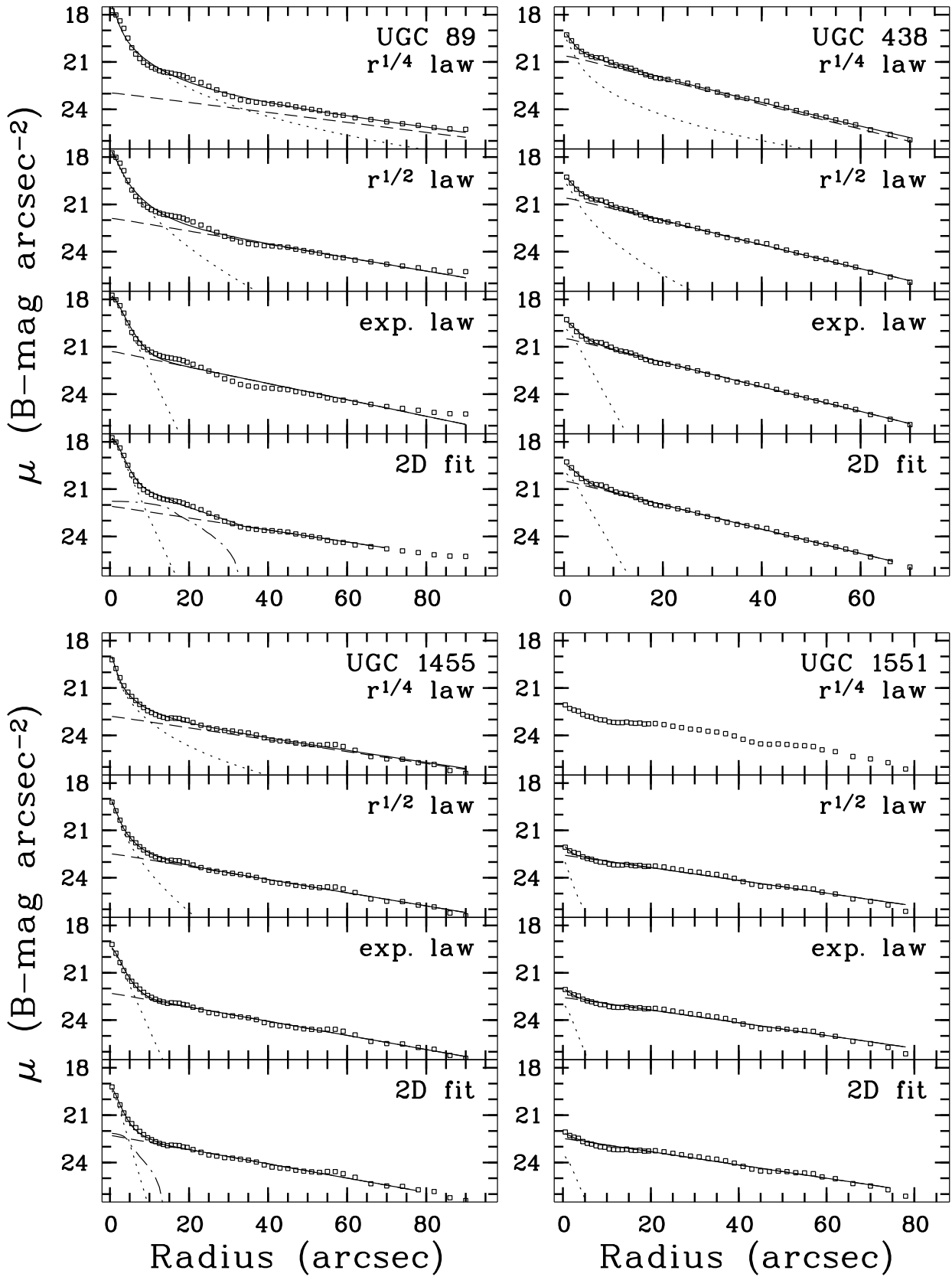}}
\caption[]{
 Some examples of the fits made to the $B$ passband surface brightness
profiles.  The squares indicate the measured profile, the dashed line
represents the fitted disk, the dotted line the fitted bulge and the
dashed-dotted line the bar.  The full line is the sum of the different
model components.  The method to extract the model luminosity profiles
from the 2D model images was the same as used on the real data. 
 \label{plfit}
 } 
 \end{figure*}

\begin{figure*}
\mbox{\epsfxsize=16.8cm \epsfbox[45 138 530 763]{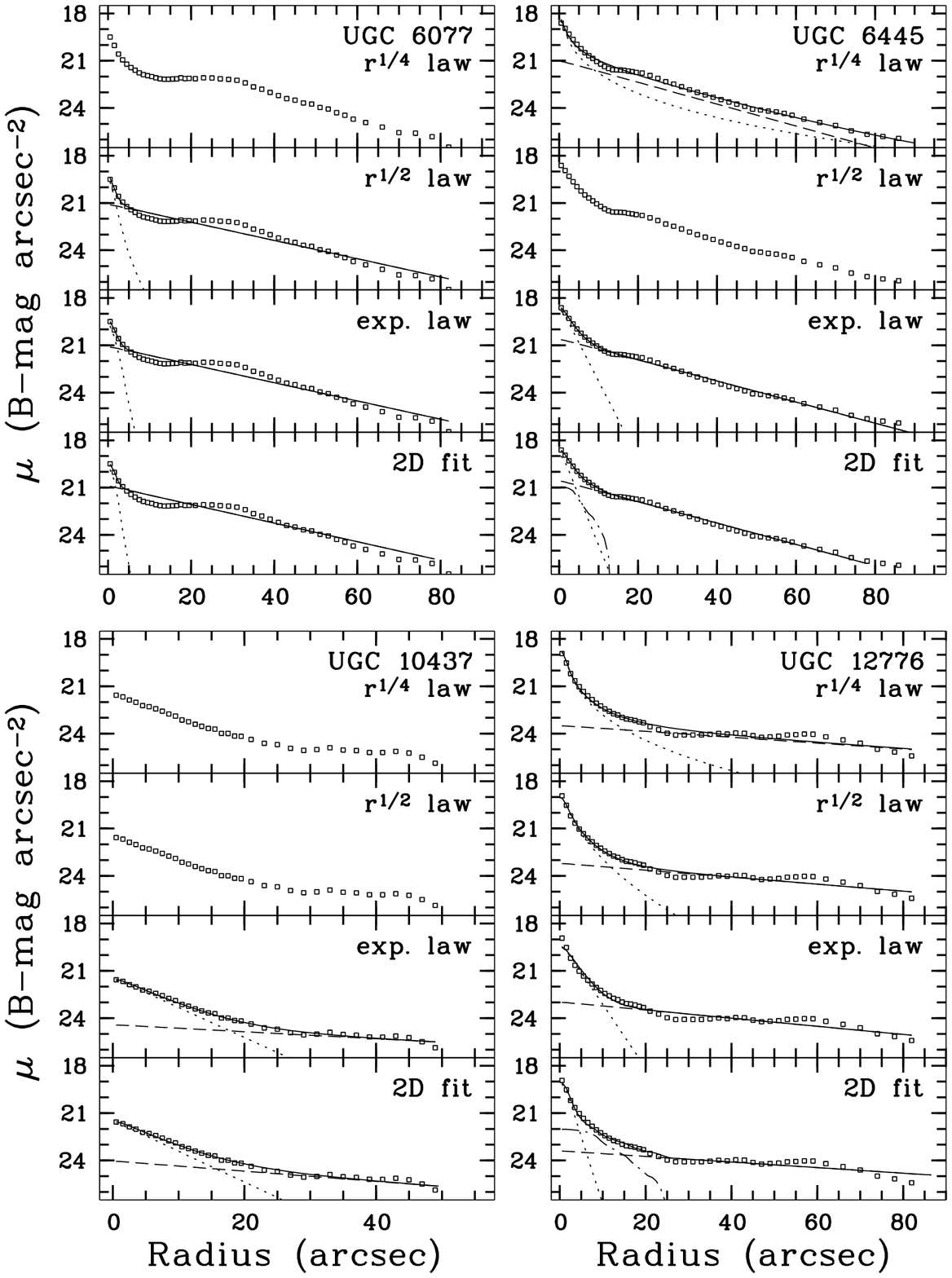}}
\ \\
\vskip 10pt
 {\small\bf Fig.~\ref{plfit}.}~-continued.\\
\end{figure*}

\begin{figure*}
 \mbox{\epsfxsize=16.7cm \epsfbox[45 140 530 763]{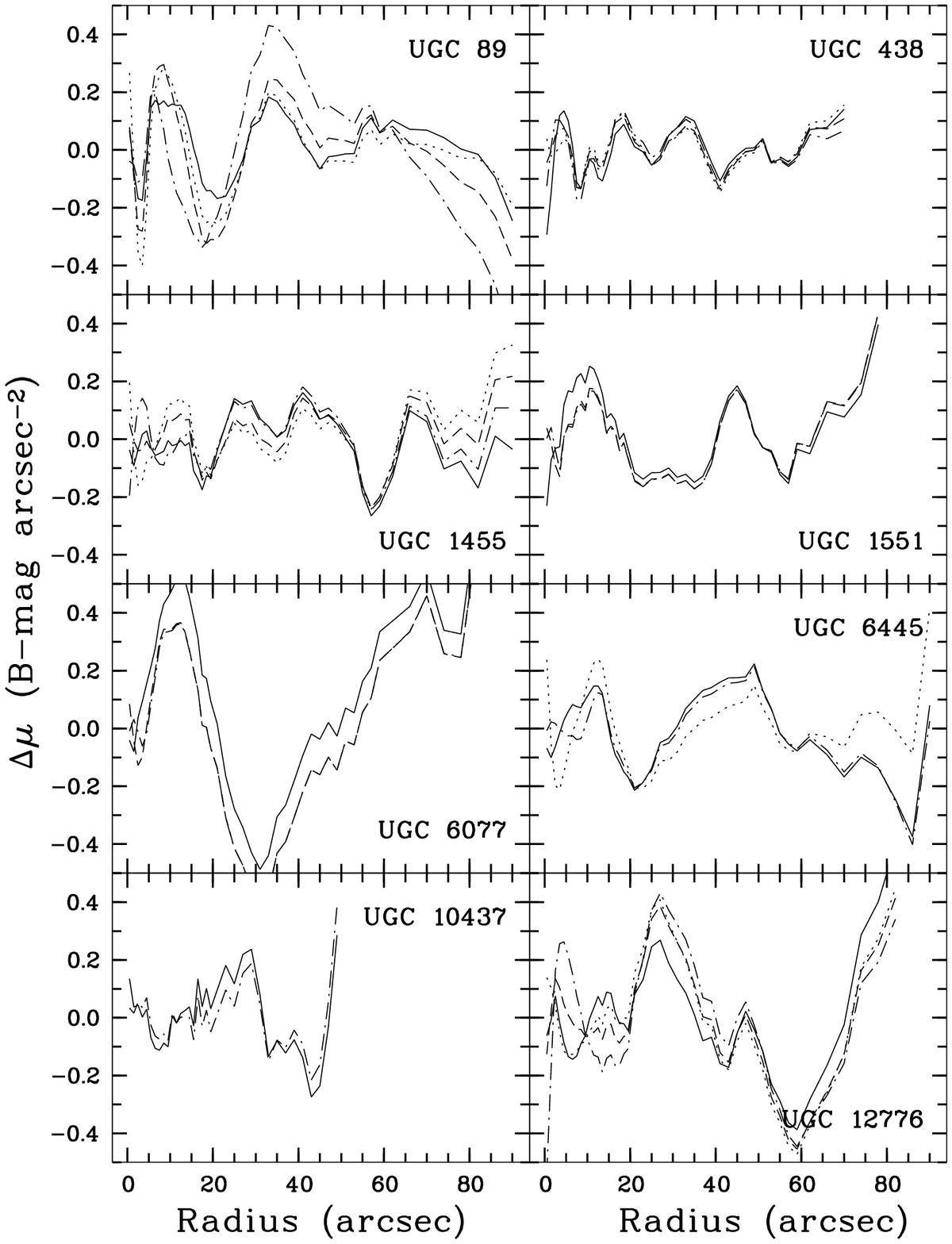}}
 \caption[]{
 The residuals $\Delta\mu$ between data and model for the galaxies
plotted in Fig.~\ref{plfit}.  The dotted line indicates the model with
the \rq law bulge, the dashed line the \rh law bulge fit, the dashed
dotted line the exponential bulge fit and the full drawn line the 2D fit
method.  Note that generally the 2D fit method results in smaller residuals,
but that the difference between the  models is small compared to the
intrinsic fluctuations in the luminosity profile.
 }
 \label{plotfitres}
 \end{figure*}

The different decompositions of eight galaxies are discussed in
more detail as it will give some insight in the problems dealing with
bulge/disk decompositions.  Figure~\ref{plfit} shows some typical best
and worst cases of four different decomposition models on $B$
passband profiles/images. Images of these galaxies can be found in
Paper~I.

 \begin{description} 
 \item[UGC\,89] is probably the clearest example in the sample of a
galaxy with three distinct components, which can only be fitted
correctly using a 2D fitting technique.  In Fig.~\ref{plfit} we see
that in the \rq law fit the bulge fills up the bar region, but also
replaces part of the disk.  The central surface brightness of the disk
is probably too faint and the scalelength too large in this fit.  When
we use an \rh or an exponential bulge, the disk tries to fill up the
bar region, making the fit to the outer exponential part of the profile
worse.  The disk parameters are obviously incorrect for these fits, but
now the surface brightness is too faint and the scalelength too small.
Only including a bar yields a satisfying result, also displayed in
Fig.~\ref{u89fit}.

\item[UGC\,438] is a galaxy with a small bulge and a distinct
exponential disk with some enhanced star formation in the spiral arms
near the center.  All fitting methods seem to be equally justified, but
notice the very extended bulge in the case of the \rq bulge profile.

\item[UGC\,1455] is a symmetrical galaxy with a small bar/oval component
in the center.  All fits seem equally justifiable as long as we assume
that the oval component is part of the bulge.  The \muo\  becomes
gradually fainter going from \rq to the exponential bulge.  If we
assume that the bar is part of the disk, which is likely as some spiral
arms start at the ends of the bar, the 2D fit is the best fit. 

\begin{figure*}
 \mbox{\epsfxsize=8.8cm\boundboxo{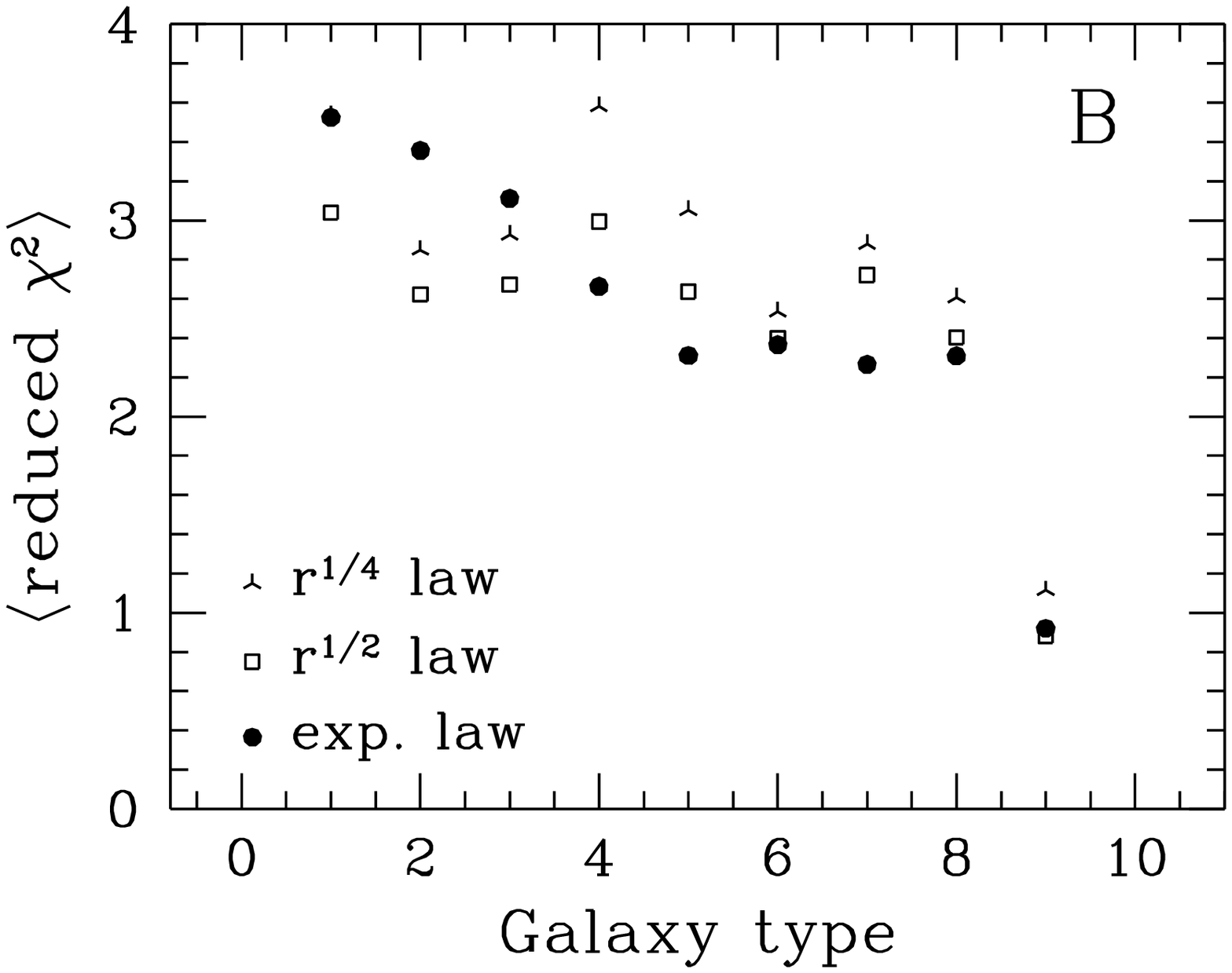}}
 \mbox{\epsfxsize=8.8cm\boundboxt{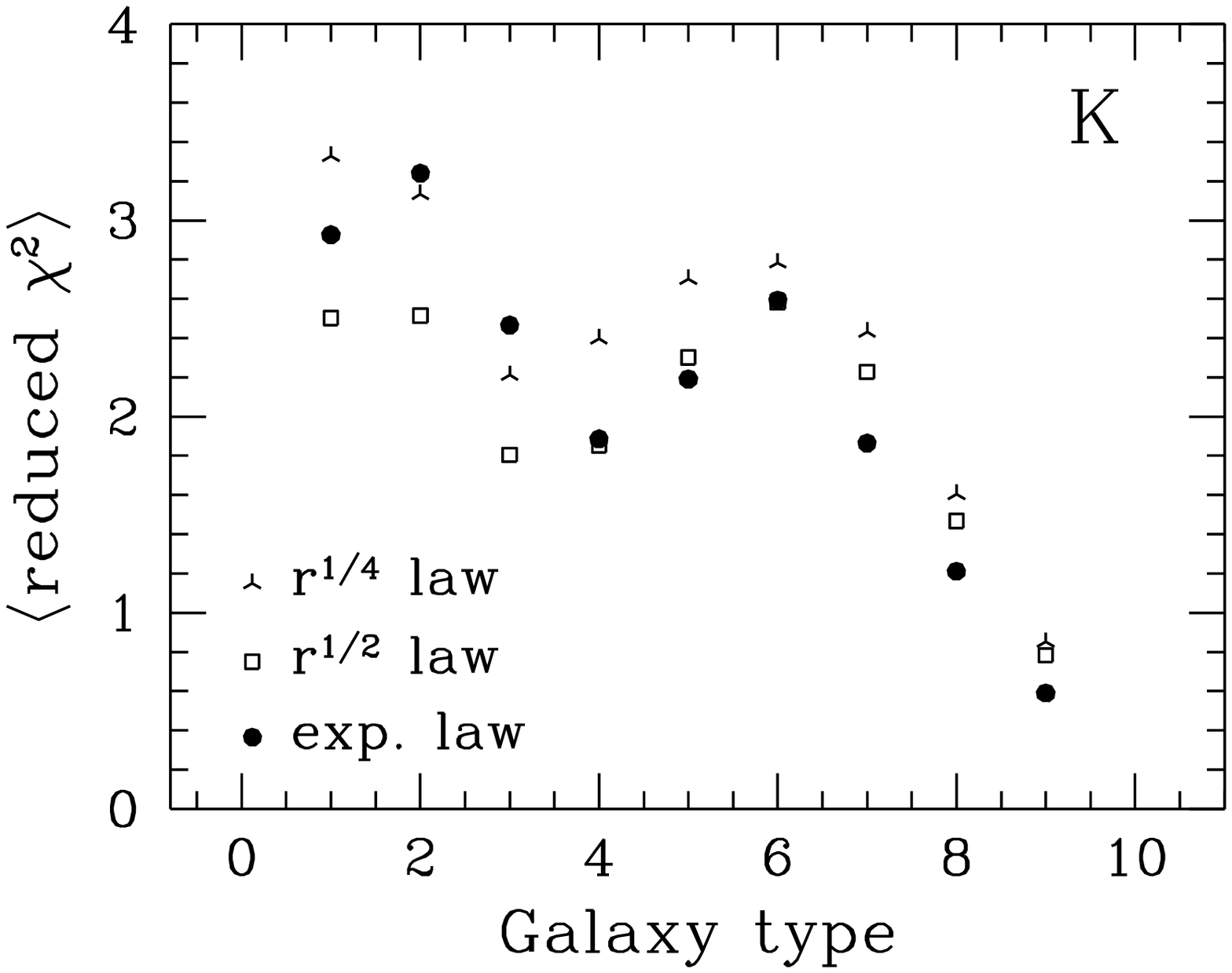}}
 \caption[]{
The average reduced $\chi^2$ values per morphological type for the 1D fit
with different bulge models.
\label{typechi2}
}
 \end{figure*}

\item[UGC\,1551] has a low surface brightness disk with some flocculent
star formation and a very small bulge.  The \rq profile did not
converge for the $B$ passband (though it did for the other passbands).
Even though convergence was reached with the other fitting functions,
the bulge results can hardly be called reliable; the small bulge is very
hard to fit.


\item[UGC\,6077] is a typical case of a Type~II profile, where the
exponential outer profile does not continue all the way inward before
the bulge takes over the luminosity in the central part of the galaxy. 
Morphologically these galaxies are in general barred with strong spiral
arms forming at the end which are tightly wound to form a ring. 
Sometimes the bar is less pronounced.  Moving to longer wavelengths the
bump becomes less pronounced as the enhanced star formation doesn't
show up so clearly at those wavelengths.  None of the tested fitting
models is able to decompose this type of galaxies well.  One has to
adjust the exponential law of the disk to contain an inner hole or one
has to model spiral arms in the disk.  None of the fitting techniques are
correct (the \rq law fit did not even converge for this galaxy) and the
central surface brightnesses and scalelengths found are just an
indication of the extent and surface brightness of the disk. 

\item[UGC\,6445], a galaxy with a small oval component, has a less
pronounced Type~II profile.  Because of a slight shift in steepness of the
luminosity profile in the outer region of the galaxy, the \rq fit will
tend to dominate in the central and outer region.  Consequently bulge
and disk are equally important over the whole extent of the galaxy in
such a fit.  Inclusion of a bar or oval component immediately
diminishes the bulge contribution at large radii, even if one would fit
an \rq bulge profile.  Clearly the 2D fit is the best option for this
galaxy. 

\item[UGC\,10437] is an example of one of the galaxies in the sample
with a very low surface brightness disk.  The fits made to this kind of
galaxy are in general very unstable as the disk disappears rapidly in
the sky noise.  This means that the profile that can be fitted only
extends over a few scalelengths.  The $H$ and $K$ passband observations
especially could pose some problems as these showed hardly anything
else other than the bulge above the sky noise (see also UGC\,334,
UGC\,628 and UGC\,9024).  The fits with \rq and \rh bulge profiles only
converged to unphysical values for this galaxy, the exponential 1D
bulge and the 2D fit yield nearly the same result, as might be expected
for a nearly face-on galaxy. 

\item[UGC\,12776] is a galaxy that consists mainly of a bar, with two
arms originating at the ends. Similar galaxies are UGC\,2368 and
UGC\,10083. Obviously a bar should be included in the models to fit
these galaxies, but we have to keep in mind that in these cases the bar
really seems to belong to the disk. The central surface brightness of
such a galaxy is probably best described by the sum of the
contributions of the disk and the bar.
 \end{description}

Figure~\ref{plotfitres} shows the residuals between models and data. 
This figure shows that including a bar in the fitting routine can indeed
improve the overall fit, but also shows that the residuals are
dominated by fluctuations in the profiles.  These fluctuations are not
due to signal-to-noise problems, but are intrinsic to the light
distribution of the galaxy. This makes it difficult to decide which
bulge model is best, as the change in reduced $\chi^2$ is small
compared to the reduced $\chi^2$ value itself.  One would need additional
components to fit the fluctuations, but this is beyond the scope of
this work.

\subsection{Comparison of $\chi^2$ values}

An exponential bulge was used in the 2D fitting technique, which was
motivated by the 1D test results.  One such test is the comparison of
the reduced $\chi^2$ values for the different bulge models. 
Figure~\ref{typechi2} shows the different reduced $\chi^2$ values as
function of morphological type.  Note that the $\chi^2$ values are
smaller in the $K$ passband than in the $B$ passband and that the
differences between the $\chi^2$ values of the different bulge models
are small compared to these $\chi^2$ values themselfs.  

{
\tabcolsep=1.67mm

\begin{table*}
\caption[]{
 The median of the fit parameter errors for all galaxies, where the
maximum error in the sky background subtraction was used to calculate
the uncertainty in the fit parameters.  The errors in \muo\ and
\mue\ are in \magarc, the relative errors in $h$ and \re\ were
defined by $\Delta h$/$h$ and $\Delta$\re/\re. 
 \label{mederr}
 }
\begin{tabular}{cccccccccccccccccc} 
\hline 
\hline 
\ \ Band \ \ \ & \multicolumn{4}{c}{median error in \muo}
&\ \ \ & \multicolumn{4}{c}{median error in $h$}
&\ \ \ & \multicolumn{3}{c}{median error in \mue}
&\ \ \ & \multicolumn{3}{c}{median error in \re} \\
      &marking&\rq &\rh &exp. 
     &&marking&\rq &\rh &exp.
     &&\rq &\rh &exp. 
     &&\rq &\rh &exp. \\
\hline
$B$ & 0.02 & 0.07 & 0.06 & 0.05 && 0.037 & 0.044 & 0.033 & 0.029 && 0.15 & 0.06 & 0.02 && 0.50 & 0.26 & 0.14\\
$V$ & 0.03 & 0.08 & 0.07 & 0.05 && 0.033 & 0.035 & 0.030 & 0.025 && 0.18 & 0.06 & 0.02 && 0.33 & 0.21 & 0.12\\
$R$ & 0.02 & 0.07 & 0.06 & 0.04 && 0.027 & 0.034 & 0.026 & 0.023 && 0.11 & 0.05 & 0.02 && 0.40 & 0.18 & 0.09\\
$I$ & 0.03 & 0.11 & 0.10 & 0.07 && 0.042 & 0.030 & 0.026 & 0.021 && 0.18 & 0.08 & 0.03 && 0.43 & 0.15 & 0.08\\  
$H$ & 0.04 & 0.16 & 0.20 & 0.12 && 0.041 & 0.039 & 0.028 & 0.023 && 0.28 & 0.12 & 0.04 && 0.43 & 0.15 & 0.09\\
$K$ & 0.04 & 0.17 & 0.14 & 0.11 && 0.049 & 0.032 & 0.025 & 0.021 && 0.27 & 0.09 & 0.04 && 0.39 & 0.16 & 0.09\\
\hline
\hline
\end{tabular}
\end{table*}

}

{
\tabcolsep=1.2mm
\begin{table*}
\caption[]
 {The mean relative change in disk parameters due to different techniques for
all galaxies where the fit routine converged to physical values for all
techniques.  All mean changes are relative to the ``marking the disk''
method, with $\Delta \mu_0$ in \magarc, $d = 2(h_{\rm mark}-h_{\rm
other})/(h_{\rm mark}+h_{\rm other})$ dimensionless.  The errors are
standard deviations. 
 }
\label{compfit}

\begin{tabular}{cr@{\ \ \ \ }rrrrr@{\ \ \ \ }rrrr}
\hline
\hline
Band& & \multicolumn{4}{c}{$\langle \Delta \mu_0 \rangle$} & & \multicolumn{4}{c}{$\langle d \rangle$}\\
   & \# & \multicolumn{1}{c}{\rq law}&\multicolumn{1}{c}{\rh law}&\multicolumn{1}{c}{exp. law}&\multicolumn{1}{c}{2D fit}& &\multicolumn{1}{c}{\rq law}&\multicolumn{1}{c}{\rh law}&\multicolumn{1}{c}{exp. law}&\multicolumn{1}{c}{2D fit} \\
\hline
$B$& 71&--0.21 $\pm$ 0.31 &--0.06 $\pm$ 0.20 &  0.06 $\pm$ 0.20 &  0.07 $\pm$ 0.24 & &--0.01 $\pm$ 0.14 &  0.02 $\pm$ 0.10 &  0.06 $\pm$ 0.11 & 0.06 $\pm$ 0.14 \\
$V$& 64&--0.21 $\pm$ 0.28 &--0.16 $\pm$ 0.53 &  0.03 $\pm$ 0.18 &  0.02 $\pm$ 0.31 & &  0.02 $\pm$ 0.12 &--0.00 $\pm$ 0.09 &  0.04 $\pm$ 0.09 & 0.03 $\pm$ 0.13 \\
$R$& 72&--0.25 $\pm$ 0.37 &--0.13 $\pm$ 0.38 &  0.05 $\pm$ 0.21 &  0.02 $\pm$ 0.30 & &--0.01 $\pm$ 0.15 &--0.01 $\pm$ 0.12 &  0.04 $\pm$ 0.11 & 0.03 $\pm$ 0.12 \\
$I$& 68&--0.28 $\pm$ 0.39 &--0.21 $\pm$ 0.50 &  0.04 $\pm$ 0.19 &  0.03 $\pm$ 0.29 & &  0.01 $\pm$ 0.15 &--0.03 $\pm$ 0.15 &  0.03 $\pm$ 0.10 & 0.03 $\pm$ 0.12 \\
$H$& 34&--0.19 $\pm$ 0.47 &--0.17 $\pm$ 0.35 &--0.00 $\pm$ 0.43 &--0.04 $\pm$ 0.56 & &  0.05 $\pm$ 0.22 &  0.00 $\pm$ 0.15 &  0.00 $\pm$ 0.09 &-0.02 $\pm$ 0.13 \\
$K$& 65&--0.22 $\pm$ 0.40 &--0.15 $\pm$ 0.45 &  0.08 $\pm$ 0.19 &  0.02 $\pm$ 0.28 & &  0.02 $\pm$ 0.18 &  0.00 $\pm$ 0.09 &  0.05 $\pm$ 0.09 & 0.02 $\pm$ 0.10 \\
\hline
\hline
\end{tabular}
\end{table*}
}

For galaxies with types later than Sb, the fits with the exponential
bulge give the smallest residuals.  For galaxies in the range from Sa-Sb,
the \rh law bulge fits give the best results.  In this range, the \rq law
and exponential law bulge give comparable results in reduced $\chi^2$ values. 
Therefore one could propose an intermediate transition type bulge for
early spirals between the \rq law ellipticals and the exponential bulges
of late-type spirals.  This could be analogous to the pure elliptical
systems, where low luminosity systems are less centrally concentrated
than high luminosity systems (Young \& Curie \cite{YouCur94}). 

The results of the study by Andredakis \& Sanders (\cite{AndSan94}) are
confirmed using this sample. Exponential bulges are statistically at
least as justified as \rq law bulges. When the galaxy image does not
contain enough information to determine the shape of the bulge, the
exponential bulge is preferred for galaxies with classification later
than Sb. For early-type spiral galaxies an \rh law bulge is a good
alternative. For consistency reasons, only the results obtained with
the exponential bulge will be used in subsequent papers when analyzing
this large sample of galaxies. 


\subsection{Comparisons of errors due to sky uncertainties}

The tests on artificial images discussed in Sect.~\ref{decomp2D}
indicated that the uncertainty in the sky background level is, once a
bulge model has been chosen, the largest source of error.  As explained
before, the maximum errors due to incorrect sky background subtraction were
only calculated for the 1D methods.  The
median relative errors for each parameter and each method are listed in
Table~\ref{mederr}.   Two trends can be seen in this table:\\
1) the median errors in the surface brightness parameters increase
going from the $B$ to the $K$ passband, while the uncertainties in the
scale parameters decrease and\\
2) the median errors in {\em all} parameters decrease moving from
the fits with the more centrally peaked \rq law bulge profile to the
fits with the shallower exponential bulge profile.

A change in sky background level affects mostly the faint outer regions
of the galaxy profiles.  The \rq law profile offers often a greater
contribution to the outer parts of galaxies than the exponential bulge
profile and will be more affected by a sky level change. The parameters
resulting from the fit with the exponential bulge are the most stable
and reproducible according to the results in Table~\ref{mederr} which
is independent of the conclusion whether this bulge model is correct or
not.

\subsection{Comparison of the resulting disk parameters}
\begin{figure}
\mbox{\epsfxsize=8.6cm\boundboxt{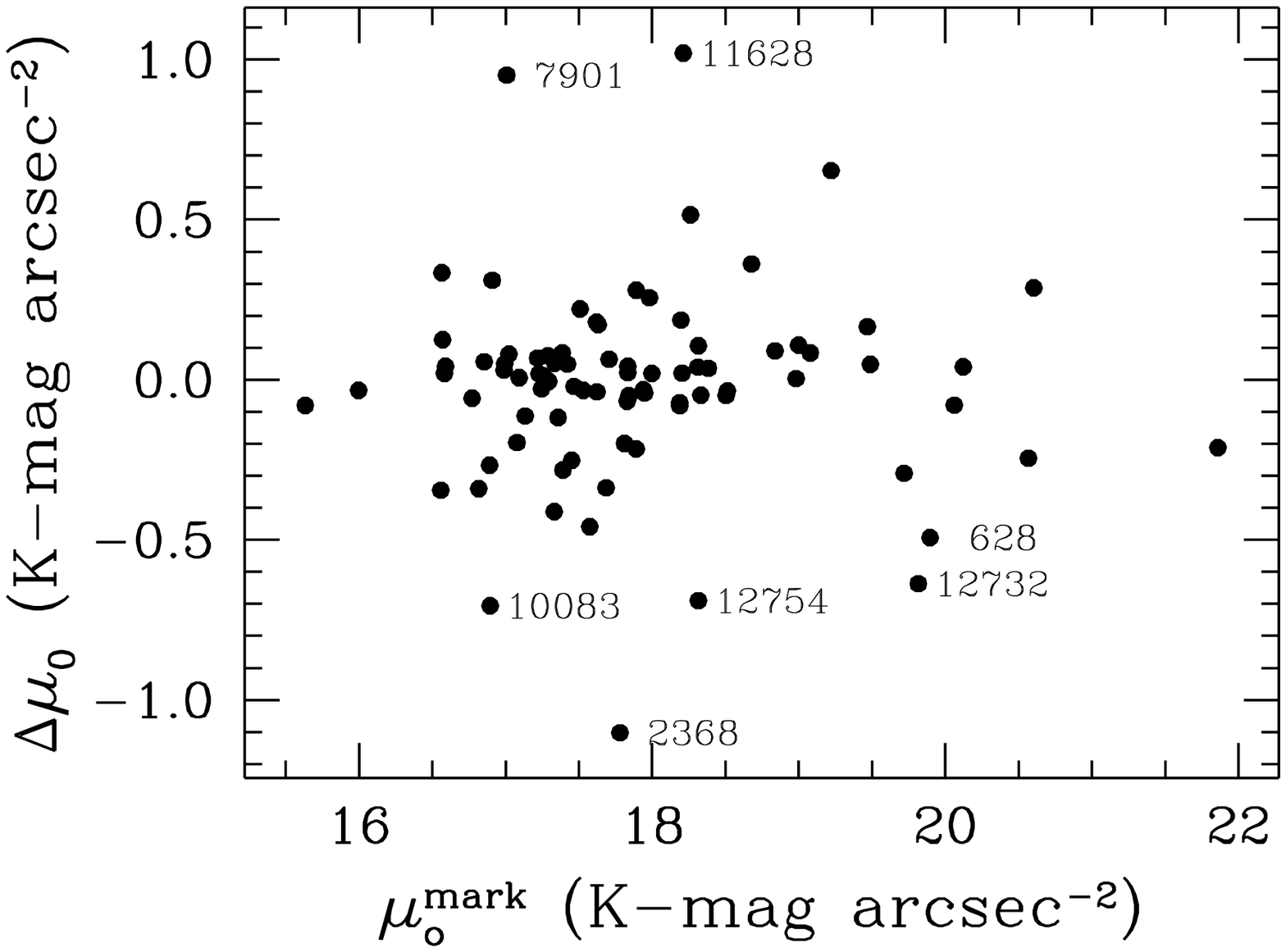}}
 \caption[]{
The difference between the central surface brightnesses obtained with the
``marking the disk'' method and the 2D fit method ($\Delta \muo =
\muo^{\rm mark} - \muo^{\rm 2D}$) as function of \muo\ in the $K$ passband.
The UGC numbers of the most deviant galaxies are indicated.
}
 \label{dcs_cs}
 \end{figure}

\begin{figure*}
\mbox{\epsfxsize=8.8cm\boundboxo{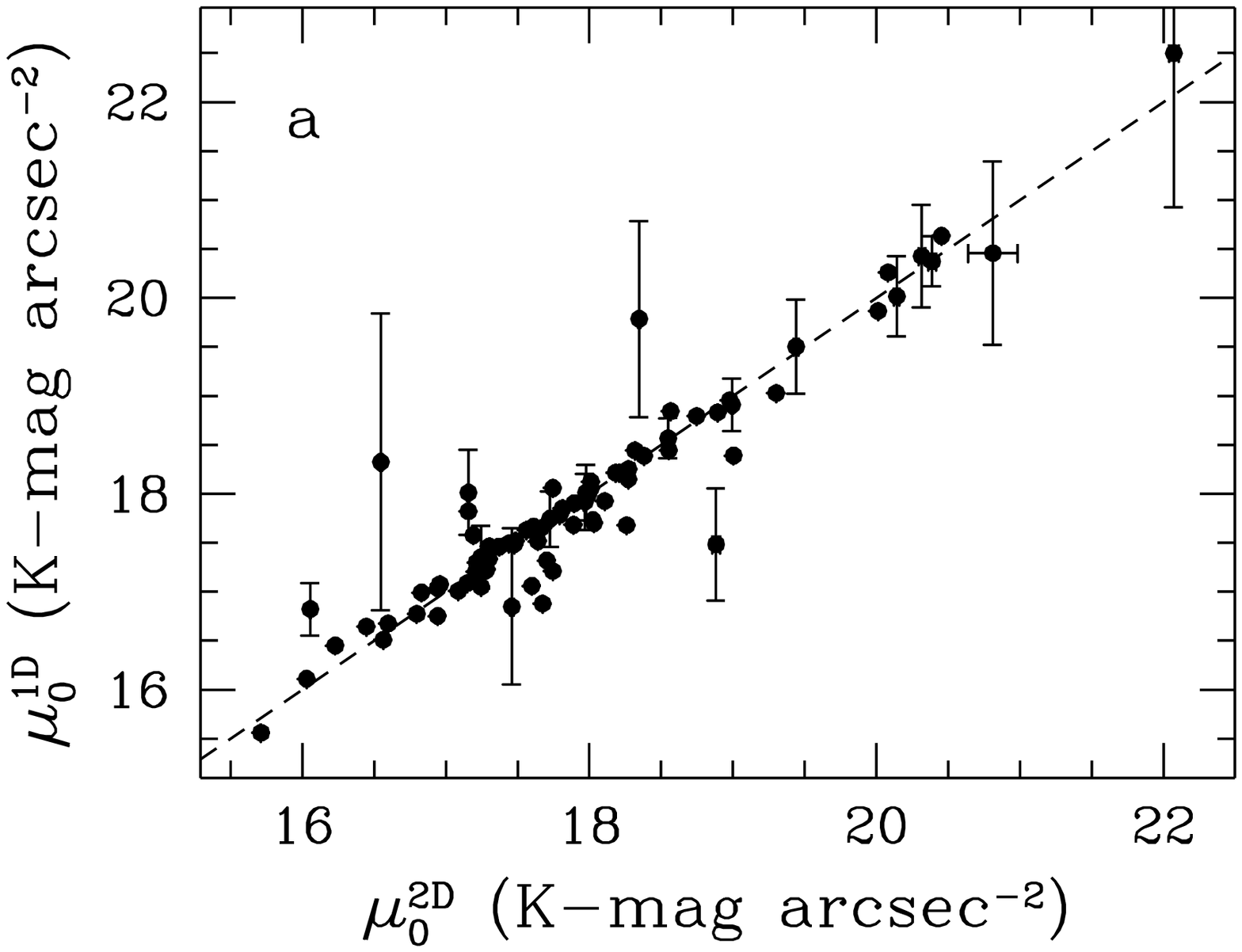}}
\mbox{\epsfxsize=8.8cm\boundboxt{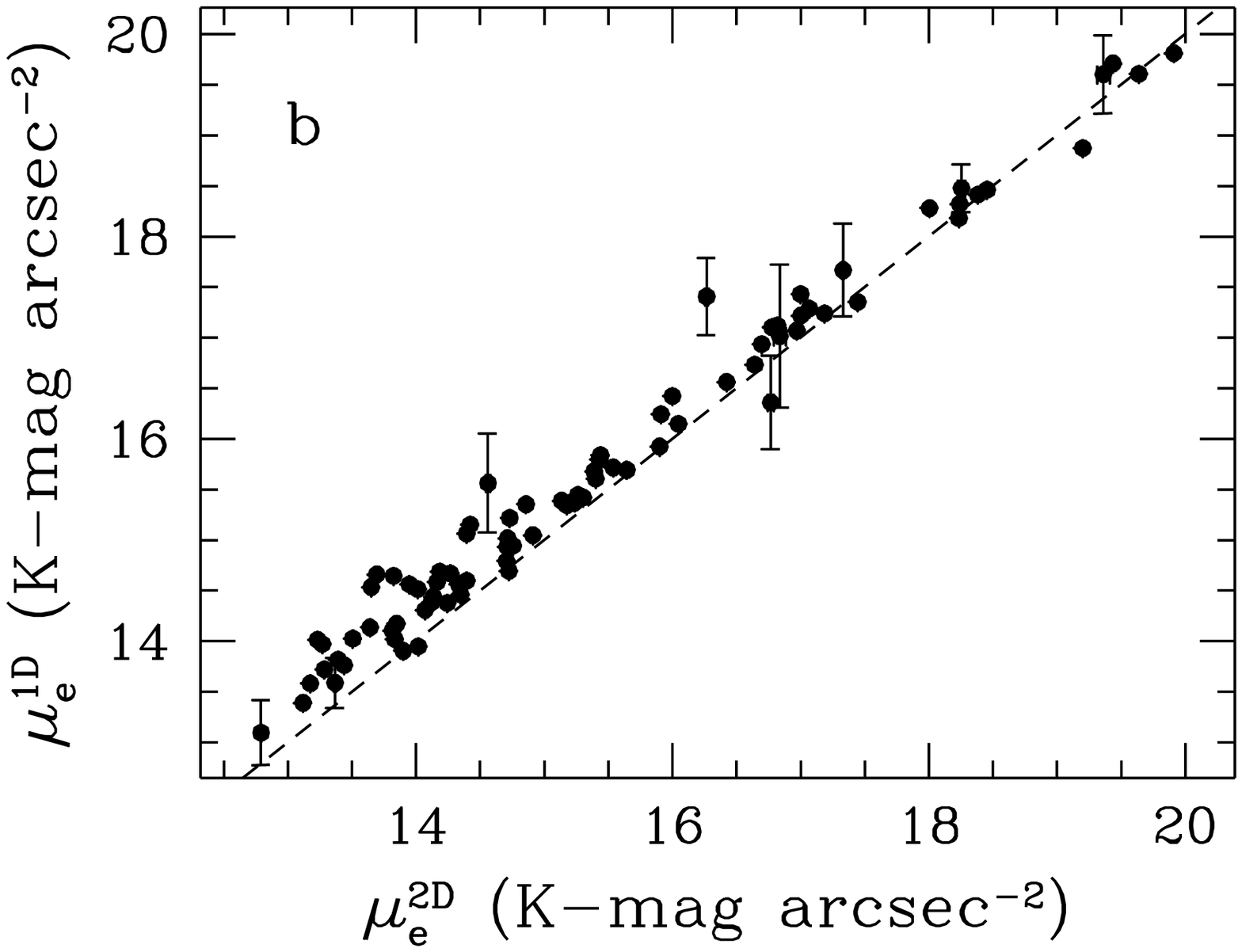}}

\vspace{0.8cm}
\mbox{\epsfxsize=8.8cm\boundboxo{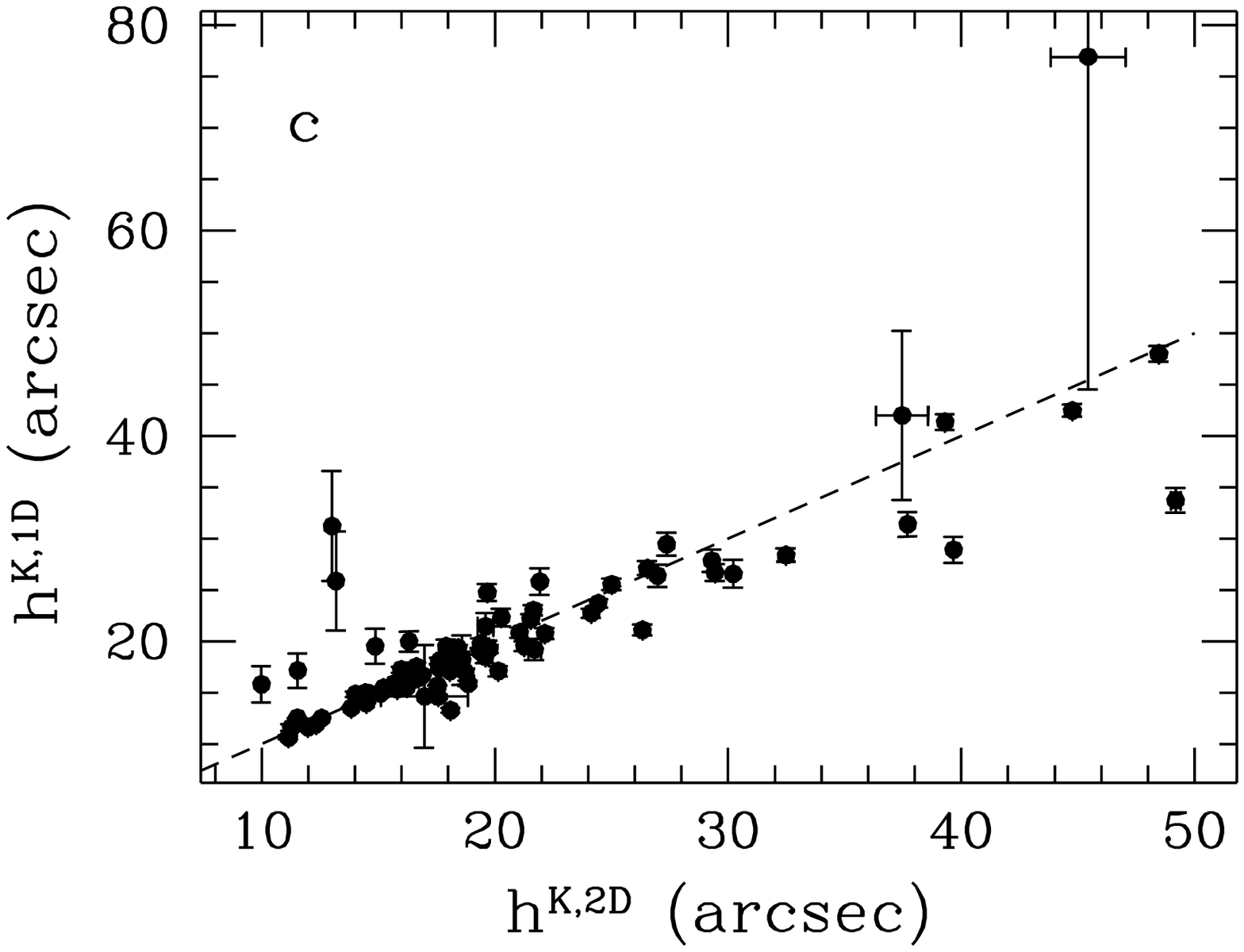}}
\mbox{\epsfxsize=8.8cm\boundboxt{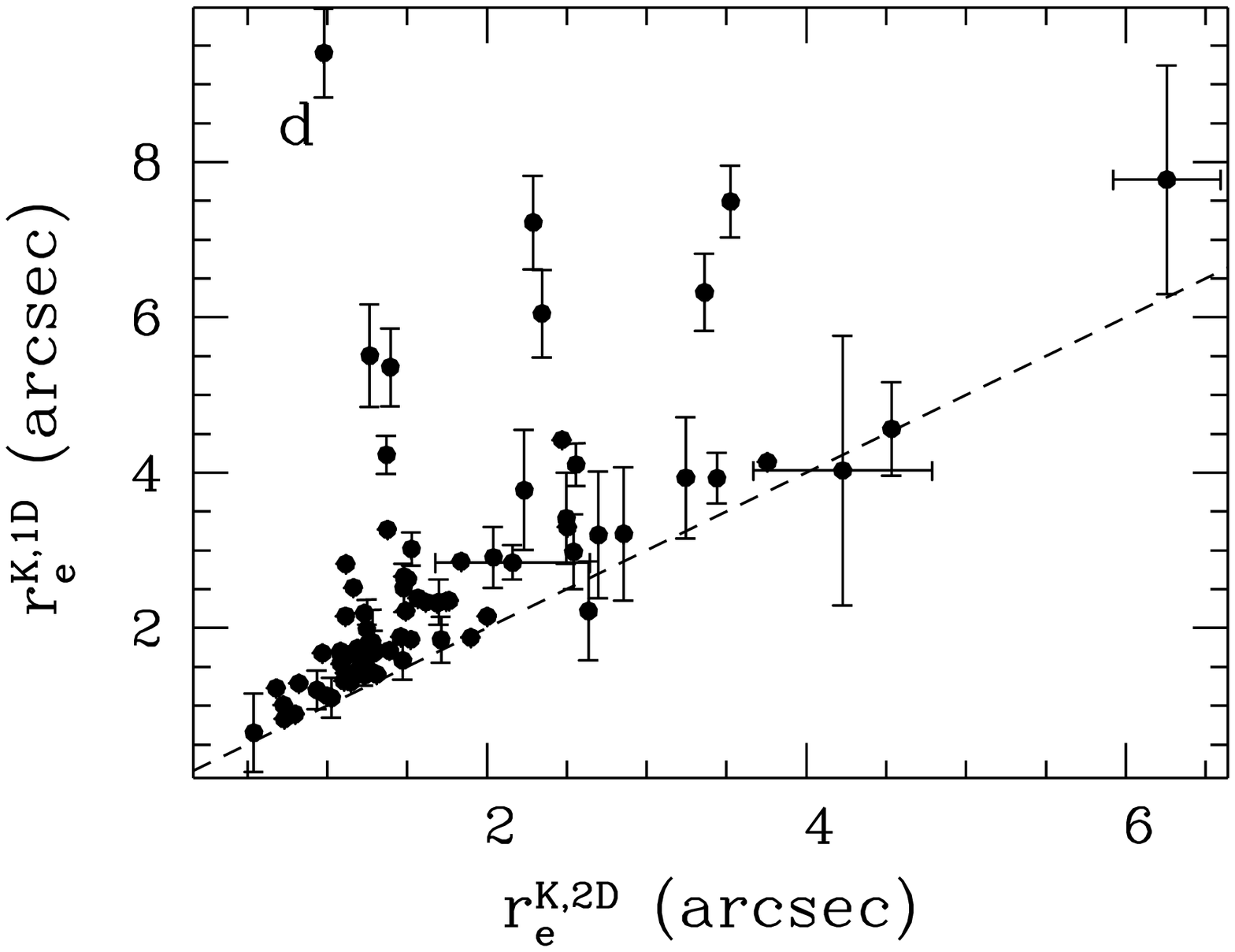}}
 \caption[]{
 A comparison between the $K$ passband parameters obtained with the 1D
and the 2D methods.  An exponential bulge profile was used for both the
1D and the 2D method.  {\bf a}) Central surface brightness of the disk, {\bf b})
effective surface brightness of the bulge, {\bf c}) scalelength of the disk
and {\bf d}) effective radius of the bulge.  The errors in the 2D parameters
are the formal fit errors of the fit routine, the errors in the 1D
parameters are the maximum errors due to the uncertainty in the sky
background subtraction. 
}
 \label{comp1D2D}
 \end{figure*}

As a final point in the comparison of the different bulge models
and fitting methods the resulting parameters are compared.
Table~\ref{compfit} lists the average differences in the disk
parameters using the different methods. The disk parameters of all
different models are compared to those of the ``marking the disk''
method.  The change in central surface brightness $\Delta \mu_0 \!= \!
\mu_{\rm 0,mark}$--$\mu_{\rm 0,other}$ is on average the largest for
the \rq law bulge and small for the exponential bulge in both the 1D
and 2D case.  The central surface brightness of the disk is lower for
the \rh and \rq law bulges as part of the disk light is replaced by the
light of these more extended bulges.  The average relative change in
scalelength ($d \!= \!2(h_{\rm mark}-h_{\rm other})/(h_{\rm mark}+h_{\rm
other})$) is a few percent at most for all models.  The scalelengths
become on average a little bit smaller.  The standard deviations of the
differences in \muo\ are much larger for the \rh and \rq law bulges
than for the exponential bulges.  Apparently the methods using
exponential bulges yield more often disk parameters that are in
accordance with the ``marking the disk'' results, which represent the
human intuition.

A comparison the results from the ``marking the disk'' fit with the 2D fit
(Fig.~\ref{dcs_cs}), shows that the central surface brightnesses agree
for most galaxies to within 0.5 mag arcsec$^{-2}$.  The few exceptions
to this can be easily explained. On the one hand, the surface
brightness is fainter in the 2D case for UGC\,2368, UGC\,10083 and
UGC\,12754, as a bar is making up a large fraction of the disk and for
some low surface brightness galaxies (UGC\,628 and UGC\,12732), because
the distinction between the nearly absent bulge and the disk is hard to
make.  On the other hand, UGC\,7901 and UGC\,11628 have a gradual change
in slope of their luminosity profile, which is interpreted differently
by the 2D fitting program than by the human eye. The 2D fit central
surface brightnesses are much brighter.

Figure~\ref{comp1D2D} shows a comparison between the 1D and 2D bulge
and disk parameters, both obtained with the exponential bulge profile.
The errorbars on the 2D results indicate the formal fitting errors of
the fitting routine, thus the uncertainty in the $\chi^2$ minimum found
by the routine. The errors in the 1D parameters result from
recalculations of the fit with the sky background uncertainties added
and subtracted. 

Two points can readily be made from these errors: 1) the
errors due to sky uncertainties are almost always much larger than the
formal fitting errors and 2) the data points which are the most deviant from
the equality line have in general also the largest uncertainty due to
the sky errors. The errors due to sky uncertainty are apparently also
indicators of the differences between both fitting methods. 

The 1D and 2D disk parameters in Fig.~\ref{comp1D2D} agree quite well
without any systematic deviations. The effective surface brightness of
the bulge shows a small but systematic deviation, the surface
brightnesses determined with the 1D method are most of the times a little
fainter. The comparison of the effective radii shows the largest
deviations. The 2D effective radii are smaller, especially for the
galaxies where a bar was fitted. The profile comparisons of
Fig.~\ref{plfit} showed that these smaller effective radii are more
realistic.

\subsection{The exponential bulge versus other bulge models}

 The exponential bulge profile was used in the 2D fitting technique
instead of the more widely used \rq law profile.  The previous tests
have shown that this is justified.  The profile decompositions look at
least as good as with other bulge models, which is reflected in the
reduced $\chi^2$ results.  Table~\ref{mederr} shows that the
exponential bulge results are least effected by the uncertainty in the
sky level and Table~\ref{compfit} shows that the exponential bulge fit
reflects the ``marking the disk'' results the most.  The choice for the
exponential bulge is obvious if one further realizes that the
exponential bulge fit, contrary to the other models, almost always
converges without fine tuning the initial parameters.  One could
propose to use \rh law bulges for early-type spiral galaxies, but for
consistency reasons I will use only the exponential bulge law results
in subsequent papers.

\section{Error discussion}
\label{discus2}

Good error estimates of the fit parameters are important to determine
the reliability of the derived relations in the next papers of this
series.  All sources of error and the typical percentage of error they
cause in the different bulge and disk parameters are discussed. 

In the decomposition methods discussed in this article one is confronted
with three kinds of uncertainties: 1) uncertainties in the component
profiles, 2) the uncertainties due to the fitting method and 3)
measurement errors.  The first two items give systematic effects in
the derived parameters, but make comparisons within such a framework
still possible.  The last item has no systematic effect, assuming
measurement errors are randomly distributed.  Each of these points
will be addressed separately, indicating the order of magnitude uncertainty
in the resulting parameters. 

\smallskip
 \noindent
{\em 1) Errors due to uncertainties in component profiles}
 \begin{itemize}
 \item The best choice for the description of the bulge light is
probably the generalized exponential (Eq.~(\ref{genexp})), as it
includes both the \rq and the exponential law.  Its mathematical
peculiarities make it here less useful in combination with an
exponential disk profile.  Therefore the exponential function for the
bulge is preferred, as it gives a first moment analysis of the part of
the bulge that is seen above the disk luminosity.  In this way, a
surface brightness and a scale size of the central region of the bulge
can be determined.  These quantities must have some physical meaning,
even though they show only a weak relation to the ones obtained with \rh
and \rq law bulges.  The average value of \muo\ of the disk will be 0.2
\magarc\ fainter using these bulges and the rms differences will
increase to 0.4 \magarc\ with respect to the ``marking the disk'' fit. 
The scalelengths of the disks are about the same for all bulge models
used.  The uncertainty in the disk parameters for a typical galaxy in
this data set is larger due to bulge profile uncertainty than due to sky
background uncertainty (compare results in Table~\ref{mederr} and
standard deviations in Table~\ref{compfit}). 

 \item The use of exponential light profiles for disks of spiral
galaxies is well established, but Type~II profiles introduce an
uncertainty.  Are we dealing with a hole in the light distribution
(intrinsic or due to dust) or do we have an extra stellar component at
intermediate radii in the form of a ring of spiral arms? Morphology and
color considerations favor the second option.  Therefore one has to
include the inner region in the fit as this is also part of the disk
and observe at wavelengths which are least affected by young stellar
populations.  In the $B$ passband, the Type~II profiles can give
uncertainties of order 20\% in the disk parameters and up to 50\% in
the bulge parameters.  In the $K$ passband the effect is strongly
reduced to order 10\% for all parameters. 
 \end{itemize}
 \noindent
 {\em 2) Uncertainties induced by different fitting methods}

 \begin{itemize}
 \item The uncertainties in the ``marking the disk'' method have been
discussed by Giovanelli et al.~(\cite{Gio94}) and by Knapen \& van der
Kruit (\cite{KnavdK}).  They showed that the uncertainties using this
method are of order 15\%.  It has been argued that this method
gives {\em systematically} wrong results (Kormendy \cite{Kor77},
Phillipps \& Disney \cite{PhiDis83}, Davies \cite{Dav90}) which will be
discussed in Paper~III (de Jong~\cite{deJ3}).  

\item The decomposition of the 1D profile in a bulge and a disk
component should in principle be better than the previous method, but
when performed with an exponential bulge profile the disk parameters
hardly change compared to the ``marking the disk'' results. 
Byun (\cite{Byun92}) showed that this method gives systematically wrong
results for inclined galaxies. 

\item The 2D fitting method is intrinsically the best method tested
here.  It includes the effect of inclination and makes it possible to
introduce a non-axisymmetric component in the form of a bar.  With
respect to the double exponential 1D fit it has little effect on the
disk parameters and on \mue\ of the bulge.  The estimate of \re\ of the
bulge improves. The tests on artificial images showed that the errors
in the parameters are at most a few percent for bulges with \re\ larger
than 1\arcsec. 

\item One source of error that has never been investigated in detail is
the choice of weighting function for the fitting routine. I have only
used weighting schemes which reduce relative errors between model and
data, but one could argue that the inner part of galaxies should get
more weight as the signal-to-noise is better in these regions. 
Likewise one could also argue that the outer part of the luminosity
profile should get more weight as many more points are sampled (this is
what happens automatically in the 2D fitting technique).  Therefore
even a single model for bulge and disk luminosity distribution has an
intrinsic uncertainty depending on where one puts the most weight for
the fitting accuracy.  It is hard to define an acceptable range of
weighting functions, but the uncertainty in the disk parameters induced
in this way is estimated to be at most 5\% for most galaxies.  It will
be of order 10\% for the bulge parameters. 
 \end{itemize}
 \noindent
 {\em 3) Uncertainties caused by measurement errors}

 \begin{itemize}
 \item The dominant source of error in both bulge and disk parameters is
the uncertainty in the sky background subtraction.  As a second order
effect $b/a$ is also important for the estimate of \muo.  The seeing
estimate influences the determination of the parameters of
relatively small bulges.  Each of these errors has at most order 10\%
effect on the disk parameters and a 20\% effect on the bulge parameters. 
The errors will be slightly larger for low surface brightness galaxies,
due to the relatively larger contribution of the sky background error. 

 \item Another source of error, independent of the fitting method
employed, stems from the determination of the zero-point of the
magnitude scale.  In Paper~I, we calculated that this was for our
sample in the range from 0.03 to 0.12 \magarc, resulting in this
contribution being of the same order of or less than the above
mentioned measurement errors. 
 \end{itemize}

\section{Conclusions}
 \label{concl2}

The bulge, disk and bar parameters of the 86 spiral galaxies presented
in Paper~I were calculated.  In the different decomposition methods
that were explored, we were confronted with three kinds of
uncertainties: 1) the uncertainties in the component profiles, 2) the
uncertainties due to the fitting method and 3) measurement errors.  The
first two items will give systematic effects in the derived parameters,
but comparisons within such a framework are still possible.  The last
item has no systematic effect, assuming measurement errors are randomly
distributed. 

In conclusion, the use of the 2D fitting technique with exponential
light profiles for both bulge and disk yields the most reproducible and
representative component parameters.  Assuming that the 2D model with
exponential profiles for both bulge and disk is a reasonable correct
description of the global structure of a galaxy, the estimated maximum
errors of the structural parameters are of order 20\%.  These errors
are dominated by the uncertainty in the sky background level.  The
errors in the disk parameters will double if the model for the bulge is
better represented by an \rq law profile.  The errors on the bulge
parameters will then be very uncertain.  The determined parameters will
be used in the subsequent articles in these series. 


The two dimensional decomposition technique exploited here has many
advantages above other often used decomposition techniques.  Still we
are far away from full scale 3D models which link complete
self-consistent dynamics of the different components to a 2D
morphological projection.

\begin{acknowledgements}
 I would like to thank Yanis Andredakis, Piet van der Kruit, Reynier
Peletier and David Sprayberry for the fruitful discussions.  Erwin de
Blok, Joyce Majewski, Ren\'e Oudmaijer and Arpad Szomoru are
acknowledged for their many useful suggestions on the manuscript. 
 I would like to thank the anonymous referee for the useful comments that
helped to improve the manuscript.
 This research was supported under grant no.~782-373-044 from the
Netherlands Foundation for Research in Astronomy (ASTRON), which
receives its funds from the Netherlands Foundation for Scientific
Research (NWO). 
 \end{acknowledgements}



\end{document}